\documentclass{emulateapj}
\usepackage{natbib}
\usepackage{color}

\newcommand{\equ}[1]{eq.~(\ref{eq:#1})}
\newcommand{\equs}[1]{eqs.~(\ref{eq:#1})}
\newcommand{\equm}[1]{(\ref{eq:#1})}
\newcommand{\Equ}[1]{Eq.~(\ref{eq:#1})}

\newcommand{\se}[1]{\S\ref{sec:#1}}
\newcommand{\fig}[1]{Fig.~\ref{fig:#1}}

\newcommand{\Fig}[1]{Figure~\ref{fig:#1}}

\newcommand{\be}{\begin{equation}}
\newcommand{\ee}{\end{equation}}
\newcommand{\bea}{\begin{eqnarray}}
\newcommand{\eea}{\end{eqnarray}}

\newcommand{\msun}{M_\odot}

\newcommand{\ifm}[1]{\relax\ifmmode#1\else$\mathsurround=0pt #1$\fi}
\newcommand{\kms}{\ifmmode\,{\rm km}\,{\rm s}^{-1}\else km$\,$s$^{-1}$\fi}

\newcommand{\Mpc}{\,{\rm Mpc}}
\newcommand{\kpc}{\,{\rm kpc}}
\newcommand{\pc}{\,{\rm pc}}
\newcommand{\Gyr}{\,{\rm Gyr}}

\newcommand{\Myr}{\,{\rm Myr}}
\newcommand{\yr}{\,{\rm yr}}

\newcommand{\ltsima}{$\; \buildrel < \over \sim \;$}
\newcommand{\lsim}{\lower.5ex\hbox{\ltsima}}
\newcommand{\gtsima}{$\; \buildrel > \over \sim \;$}
\newcommand{\gsim}{\lower.5ex\hbox{\gtsima}}
\newcommand{\prop}{\propto}

\def\secpush{\vskip0pt plus 2.0\baselineskip\penalty-250
             \vskip0pt plus -2.0\baselineskip }

\def\sy{\,M_\odot\, {\rm yr}^{-1}}
\def\Mpcc{\,{\rm Mpc}^{-3}}

\def\omm{\Omega_{\rm m}}
\def\oml{\Omega_{\Lambda}}

\def\Mv{M_{\rm v}}

\def\Rv{R_{\rm v}}
\def\Vv{V_{\rm v}}

\def\Mg{M_{\rm g}}

\def\Msy{M_\odot {\rm yr}^{-1}}
\def\Rc{R_{\rm c}}
\def\Rd{R_{\rm d}}
\def\Re{R_{\rm eddy}}
\def\Mc{M_{\rm c}}
\def\Md{M_{\rm d}}
\def\Ms{M_{\rm sph}}
\def\Mb{M_{\rm bar}}
\def\Mt{M_{\rm tot}}
\def\Mst{M_{\rm *}}

\def\fb{f_{\rm b}}
\def\fg{f_{\rm g}}
\def\fss{\delta_{\rm ss}}
\def\fs{\delta_{\rm stab}}
\def\f{\delta}
\def\gamc{\gamma_{\rm c}}
\def\Vin{V_{\rm in}}
\def\vin{v}

\def\th{t_{\rm H}}
\def\tv{t_{\rm v}}
\def\td{t_{\rm d}}
\def\te{t_{\rm enc}}
\def\tm{t_{\rm mig}}
\def\tevac{t_{\rm evac}}
\def\ta{t_{\rm acc}}

\def\ts{t_{\rm stab}}
\def\tst{t_{\rm *}}
\def\tdis{t_{\rm dis}}
\def\tcool{t_{\rm cool}}
\def\tc{t_{\rm c}}

\def\Mdd{\dot{M}_{\rm d}}
\def\Msd{\dot{M}_{\rm sph}}
\def\Med{\dot{M}_{\rm evac}}
\def\Mbd{\dot{M}_{\rm bar}}
\def\Mstdc{\dot{M}_{*c}}
\def\Mstdd{\dot{M}_{*d}}
\def\Mstd{\dot{M}_*}
\def\Mdshear{\dot{M}_{\rm shear}}
\def\tshear{t_{\rm shear}}
\def\Nj{N_{\rm J}}

\def\Qc{Q_{\rm c}}
\def\Qs{Q_{\rm *}}
\def\Qg{Q_{\rm g}}
\def\sigc{\sigma_{\rm c}}
\def\sigs{\sigma_{\rm *}}
\def\sigg{\sigma_{\rm g}}
\def\Sigs{\Sigma_{\rm *}}
\def\Sigg{\Sigma_{\rm g}}
\def\siggs{\sigma_{\rm g*}}

\shorttitle{Massive Galaxies at High Redshift: Streams, Clumps \& Spheroids}
\shortauthors{Dekel, Sari and Ceverino}

\begin{document}

\title{Formation of Massive Galaxies at High Redshift:\\ 
Cold Streams, Clumpy Disks, and Compact Spheroids}

\author{Avishai Dekel, Re'em Sari and Daniel Ceverino}
\affil{Racah Institute of Physics, The Hebrew University, Jerusalem 91904 
        Israel}
\email{dekel@phys.huji.ac.il}
\email{sari@phys.huji.ac.il}
\email{ceverino@phys.huji.ac.il}






\begin{abstract}
We present a simple theoretical framework for massive galaxies at high
redshift, where the main assembly and star formation occurred, and report on
the first cosmological simulations that reveal clumpy disks consistent with our
analysis.  The evolution is governed by the interplay between smooth and clumpy
cold streams, disk instability, and bulge formation. Intense, relatively smooth
streams maintain an unstable dense gas-rich disk. Instability with high
turbulence and giant clumps, each a few percent of the disk mass, is
self-regulated by gravitational interactions within the disk. The clumps
migrate into a bulge in $\lsim\!10$ dynamical times, or $\lsim\!0.5\Gyr$.
The cosmological streams replenish the draining disk and prolong the clumpy
phase to several Gigayears in a {\it steady state}, with comparable masses in 
disk, bulge and dark matter within the disk radius.  The clumps form stars in 
dense subclumps following the overall accretion rate, $\sim100\sy$, and each 
clump converts into stars in $\sim0.5\Gyr$.  While the clumps coalesce 
dissipatively to a compact bulge, the star-forming disk is extended because 
the incoming streams keep the outer disk dense and susceptible to instability
and because of angular momentum transport.  
Passive spheroid-dominated galaxies form when the streams are more clumpy: 
the external clumps merge into a massive bulge and
stir up disk turbulence that stabilize the disk and suppress in situ clump and
star formation. We predict a bimodality in galaxy type by $z\sim3$, involving 
giant-clump star-forming disks and spheroid-dominated galaxies of suppressed  
star formation. After $z \sim 1$, the disks tend to be stabilized by the 
dominant stellar disks and bulges. Most of the high-$z$ massive disks are 
likely to end up as today's early-type galaxies.
\end{abstract}


\section{Introduction}
\label{sec:intro}

The common picture of galaxy formation assumes that disk galaxies form
by gas accretion and they transform into spheroidal stellar systems 
mostly by major mergers \citep{toomre72,wr78,fall79,blum84,mmw98}.
The disks are assumed to be the sites of quiescent star formation while the
mergers are responsible for intense starbursts as well as subsequent
quenching of star formation in the spheroidal merger remnant 
\citep[e.g.,][]{hopkins07}.
While major mergers do occur, recent high-redshift observations and 
theoretical developments indicate that this is not the major mode of 
galaxy formation and the star formation in them (see below).

It appears that the most effective star formers in the Universe were galaxies
of baryonic mass $\sim\!10^{11}\msun$ in the redshift range 
$z\!\simeq\!1.5\!-\!3$. 
The typical cases are represented by 
UV-selected galaxies termed BX/BM \citep{adelberger04}
and rest-frame optically selected galaxies termed sBzK \citep{daddi04},
which we refer to collectively as massive ``Star-Forming Galaxies" (SFGs). 
Their mean comoving space density is $n\!\simeq\!2\!\times\!10^{-4}\Mpcc$
\citep{tacconi08},
implying within the standard $\Lambda$CDM cosmology that they reside in
dark-matter halos of masses $\Mv \leq\!3.5\!\times\!10^{12}\msun$.
The SFGs show star-formation rates of order $100\sy$ 
\citep{genzel06,forster06,elmegreen07,genzel08,stark08},
much higher than the few $\sy$ in today's Milky Way, while their baryonic 
masses and dynamical times are comparable to the Milky Way's.
However, as argued below,
their observed properties are incompatible with being ongoing mergers 
\citep{shapiro08} or merger remnants \citep{bournaud09}.
The kinematics of many of the SFGs is consistent with a rotating disk, 
of circular velocity $V \sim 200\kms$ and high velocity dispersion 
$\sigma \sim 50 \kms$.
The typical morphology is of a thick, gas-rich disk, extending to a radius
$\Rd \sim 10\kpc$. A unique feature of these galaxies, distinguishing them
from low-redshift galaxies, is that the disk tends to be broken into several 
giant clumps, of $\sim 1 \kpc$ in size and up to a few times 
$10^{9}\msun$ each, in which most of the star formation occurs 
\citep{vandenbergh96,elmegreen04,elmegreen05,forster06,genzel08}. 
This is why they are sometimes referred to as ``chain" or ``clump cluster" 
galaxies.
There is evidence that these clumps have formed internally in the disk
\citep{bournaud08,vanstarkenburg08,shapiro08}.
In many cases there is also a significant central bulge, of a somewhat 
older stellar population than in the disk clumps \citep{genzel08},
but somewhat younger than in regular spirals \citep{elmegreen08a}.

In contrast, an ongoing merger is not expected to show the kinematics and
morphology of a rotating disk. The enhanced SFR in a merger is expected 
to occur either in the two progenitor centers or in the compact coalescing
core --- not in the outer parts of an extended disk. Merger simulations show
that a very wet merger with a special choice of
orbital parameters may leave behind a 
non-negligible disk, but in most cases it is embedded in a more massive stellar 
spheroid remnant 
\citep{springel05,robertson06_disk,robertson08,governato09,bournaud09}.
These disks are expected to be smooth and not fragmented into giant clumps 
\citep{robertson08,bournaud09}.
Independently, the space number density of SFGs is higher by about a factor of 
$4$ than the expected density of merger-induced starbursts 
of comparable SFR \citep{elmegreen07,jogee08,dekel09}.
It should be noted that the most extreme star formers at high redshift
are the dusty, bright Sub-Millimeter Galaxies (SMGs)
\citep{chapman04,tacconi08,wall08},
with SFRs 
of hundreds of $\sy$, 
but these are rarer cases, with 
space densities that are lower by an order of magnitude.
While a significant fraction of the SMGs could indeed be starbursts induced by 
major mergers, the much more common SFGs pose the interesting question
of how could such massive galaxies form most of their stars so rapidly 
at early times and through a process other than a major merger.
A necessary condition is a steady intense gas supply.

In parallel to the existence of SFGs, observations indicate that the other half 
of the galaxies of $\sim 10^{11}\msun$ at $z \sim 2$ are actually compact 
spheroids of radii $\sim 1 \kpc$ in which the SFR is suppressed to below 
$\sim 10 \sy$ \citep{kriek06,dokkum08}. 
It seems that a division of galaxy type into a Blue Cloud and a Red 
Sequence, reminiscent of the robust bimodality known from lower redshifts 
\citep[a review in][]{db06}, is already established by $z \sim 2$.
A well developed Red Sequence at $z \sim 2$ 
is not reproduced by current semi-analytic models of galaxy formation,
and cannot be explained by the infrequent major mergers.
The open questions are how do so many massive spheroids form so early, 
why are they so compact, and what is the mechanism responsible for the
suppression of star formation in these galaxies. 

The gravitational fragmentation of gas-rich, thick turbulent disks into 
big clumps,
and the subsequent migration into a central bulge, have been proposed 
\citep{vandenbergh96,elmegreen05,genzel08,bournaud08} and
successfully simulated for idealized disks in isolation 
\citep{noguchi99,immeli04_a,immeli04_b,bournaud07c}.
According to \citet{toomre64}, a disk becomes unstable once its surface density 
is high enough and the circular velocity and velocity dispersion
are sufficiently low. The former is responsible for the local self-gravity 
and the latter tend to balance gravity against local collapse.
Being gaseous helps maintaining a disk configuration
because the gas cools while a stellar disk tends to acquire and maintain a high
velocity dispersion and thicken with time. 
If the disk is marginally unstable, and if the velocity dispersion to rotation
ratio is high, which 
is equivalent to having a high disk mass compared to the total mass within the
disk radius, the typical clumps are big. 
Being massive, the timescale for clump migration
into the center, by clump-clump gravitational interactions and by 
dynamical friction, is short. 
The isolated simulations showed that the clumps coalesce into a central 
``classical" 
bulge \citep{elmegreen08b} in several disk dynamical times, 
on the order of a few hundred $\Myr$, leaving behind a stable 
low-surface-density disk.  
  
However, we deduce from 
the appearance of giant clumps in a significant fraction of the massive 
galaxies at $z \sim 2$ \citep[e.g.,][]{elmegreen07,tacconi08}
that in each of these galaxies
the clumpy disk configuration should last at least for a period comparable to 
the age of the Universe at that epoch, $\sim 3 \Gyr$,
rather than be a single short episode of migration.
The key for making it a long-term phenomenon is, again, a continuous, 
rapid supply of smooth cold gas that would replenish the disk as it is 
being drained.  The incoming gas should keep the disk at the high density that  
ensures instability and new clump formation in the disk
as the older clumps migrate inward.
Indeed, such a continuous gas supply is very natural when the galaxy is
considered in a cosmological context at high redshift. 
Theoretical work \citep{bd03,db06}, and hydrodynamical cosmological simulations
\citep{keres05,db06,ocvirk08,dekel09},
have demonstrated that the galaxies in dark-matter halos of 
$\sim 10^{12}\msun$ at $z \sim 2$ are typically Stream-Fed Galaxies, which 
are built by a few cold narrow streams at rates $\sim 100 \sy$. 
\citet{dekel09} have found that,
on average, two thirds of the stream mass are in rather smooth flows 
involving only clumps smaller than $\sim 10^{10}\msun$,
leading to mini-minor mergers of mass ratio below 1:10 relative to the 
existing galaxy. 
This has been shown to explain the abundance of high-redshift SFGs  
forming stars at high rates, three-quarters of which not through mergers.
We show below that the rather smooth streams allow the continuous 
replenishment of a disk with high surface density without over-producing 
a velocity dispersion that could have stabilized the disk or completely 
destroyed it.

On the other hand, we show that the clumpy component of the incoming streams 
does stir up the disk and can create a high velocity dispersion.
Combined with growing a massive bulge, this may  
stabilize the disk against the formation of in situ giant clumps.
If the high SFR is in the disk clumps, the elimination of these clumps
is a quenching mechanism that suppresses the star formation
preferentially in galaxies that are fed by clumpy streams
and have a high bulge-to-disk ratio.

We thus propose a scenario where the formation of galaxies
is driven by cold streams, violent disk instability and 
the growth of a spheroid. 
The cosmological cold streams play a major role in this process.
When their smooth component dominates, they build up the disks and make
them fragment into giant clumps. These in situ clumps give rise to rapid star 
formation as well as to the formation of compact bulges. In the cases of
clump-rich incoming streams, 
the external clumps help growing a massive spheroid 
through mergers and thus conspire with this spheroid to stabilize the 
disk and quench star formation.
We argue that these are the two main modes of galaxy formation,
dominating the buildup of massive galaxies at high redshift, where most
of the mass has assembled into galaxies and most of the stars have formed.

In this paper we present the theoretical framework for this scenario,
via simple analytic estimates of characteristic timescales. 
In \se{clumps} we formulate the basic quantities relevant for giant-clump
disk instability in terms of the key parameter $\f$, 
the ratio of disk to total mass within the disk radius.
In \se{dispersion} we address the self-regulation of the disk instability
by the in situ clumps themselves.
In \se{mig} we estimate the timescales for clump migration and disk evacuation.
In \se{accretion} we discuss the cosmological gas input rate through streams,
smooth and clumpy.
In \se{SS} we work out the evolution toward a steady state with in situ
giant clumps when fed by rather smooth streams.
In \se{stab} we address the effect of clump-rich streams in
growing a massive spheroid and joining forces with it toward
stabilizing the disk and suppressing the SFR.
In \se{SFR} we refer to the SFR within the disk clumps.
In \se{survive} we address the survivability of the in situ clumps,
against rapid collapse and against disruption by stellar feedback.
In \se{size} we explain the spatial extent of the clumpy disks and the 
compactness of the spheroids.
In \se{stellar} we address the instability of a combined disk if gas and stars
and the difference between the disks at high redshift and at low redshift.
In \se{simu} we display preliminary maps from the first
cosmological simulations that reveal disks in a steady-state giant-clump phase.
These high-resolution simulations are analyzed in a companion paper
\citep{ceverino09}.
In \se{conc} we summarize our analysis and discuss its implications.

\secpush
\section{Disk Instability: Giant Clumps}
\label{sec:clumps}

\subsection{Instability Criterion}

According to the standard Toomre instability 
analysis \citep[][Chapter 6]{toomre64,bt08},
a thin rotating gaseous disk becomes unstable 
to axisymmetric modes 
once the local gravity overcomes both differential rotation and pressure
due to turbulence or thermal motions. 
This is expressed in terms of the stability parameter $Q$ being 
smaller than a critical value of order unity, 
\be
Q=\frac{\sigma_r \kappa}{\pi G \Sigma} < \Qc\, .
\label{eq:Q0}
\ee
Here $\sigma_r$ is the radial velocity dispersion in the disk 
(or the gas sound speed if it is larger),  
$\Sigma$ is the surface density of the disk,  
and $\kappa$ is the epicyclic frequency.
The latter is related to $\Omega$, the angular circular velocity at radius 
$r$, by $ \kappa^2 = r\, {d\Omega^2}/{dr} +4\Omega^2$.
The value of $\kappa$ ranges from $\Omega$ for Keplerian orbits,
through $\sqrt{2}\Omega$ for a flat rotation curve, 
$\sqrt{3}\Omega$ in 
a uniform disk, to 
$2\Omega$ for solid-body rotation.  
We adopt hereafter $\kappa=\sqrt{3}\Omega$, 
appropriate for high-redshift disks. 
Note that $\Omega(r)$ is determined by the total gravitational force at $r$,
which can be partly exerted by the disk itself and partly by a more spheroidal
mass component.
A value of $Q$ below unity guarantees that there are scales of perturbations
that are both Jeans unstable and rotation unstable, namely, they are 
large enough not to be stabilized by pressure 
($\lambda>\lambda_{\rm J} = \sigma_r^2/G\Sigma$), 
and small enough not to be stabilized by the centrifugal force acting in 
the 
frame of the local perturbation
due to the disk rotation 
($\lambda <\lambda_{\rm rot} = 4\pi^2 G\Sigma/\kappa^2$ 
for a cold disk). 

For a {\it thick disk}, the analysis of radial modes is qualitatively similar, 
as long as the perturbation length scale is smaller than the
radius of the disk 
and larger than its thickness. It yields a slightly smaller value of the 
critical value, $\Qc \simeq 0.68$, for an isothermal thick disk
\citep{goldreich65_thick}.
The pressure in the high-redshift thick disks, where $\sigma_z \sim 50 \kms$,
is clearly dominated by macroscopic motions, which we crudely refer to as 
``turbulence".

For a {\it stellar disk}, 
the instability criterion is similar to a gaseous disk,
except that the factor $\pi$ in the definition of $Q$ is
replaced by 3.36 --- a negligible difference compared to the other
uncertainties in the analysis. 
Thus, as long as the velocity dispersions of gas and stars are comparable,
the instability analysis is valid for the combined system of gas and stars
as a whole. 
We adopt this approximation for our simplified analysis of the high-redshift, 
gas-dominated disks.  Observational evidence for this comes so far only from 
the thickness of the old stellar disk in ``chain" galaxies as indicated
by IR measurements, which is comparable to the thickness in the UV, 
associated with stars in formation and therefore gas \citep{elmegreen09}.  
In \se{stellar}, we address deviations from this simple case, where we
discuss star-dominated disks at lower redshifts.

\subsection{Clumps and Transient Features}

The characteristic wavelength of the unstable mode is
\citep[e.g.,][Fig.~6.13]{bt08}
\be
\lambda_c \simeq \frac{2\pi^2G\Sigma}{\kappa^2} \ .
\label{eq:lambdac}
\ee
This is $\simeq 0.5\lambda_{\rm rot}$ 
and $\simeq 2 \lambda_{\rm J}$ for $Q\simeq 1$.
The ring-like density perturbation is expected to break into lumps, 
and we take the pre-collapse radius of a typical lump 
to be $\Rc \simeq \lambda_{\rm c}/4$.

If the timescale for decay of pressure support in the growing perturbations
is comparable to or shorter than the disk crossing time $\Omega^{-1}$, the 
perturbations become gravitationally bound virialized clumps of 
radii $\lsim \Rc$. 
If the pressure decay time is longer, the perturbations are stretched by 
shear on a dynamical timescale and become elongated transient features on 
a length scale comparable to the disk scale \citep{goldreich65_shear}.
For example,
\citet{gammie01} simulated the case of a thin Keplerian gaseous disk 
and found that a cooling time as short as $\tcool < 3 \Omega^{-1}$
permits the fragmentation into bound clumps.
He demonstrated that for $\tcool \gsim 3 \Omega^{-1}$, 
the system reaches a gravo-turbulent steady state dominated by transient
features in which the cooling is balanced by the dissipation of turbulence,
by shocks or via a turbulent cascade to the viscous scale.
In our case of a gas-dominated high-$\sigma$ galactic disk, 
the gas radiative cooling time is much shorter than the dynamical
time\footnote{The radiative cooling time is 
$\tcool \simeq 2.6\times 10^{3} n^{-1} T_4 \Lambda_{-22}^{-1} \yr$,
where $n$ is the gas density in atoms per ${\rm cm}^3$, here $n \sim 10$, 
$T_4$ is the temperature in $10^4$K,
and $\Lambda_{-22}(T)$ is the cooling rate in units of 
$10^{-22} {\rm K}\,{\rm cm}^3\,{\rm s}^{-1}$.
At $T_4 \gsim 1.8$, $\Lambda_{22} \sim 1$,
and at lower temperatures it drops sharply: at $T_4=1$, $\Lambda_{22} \sim
0.04$, and below it is roughly $\Lambda \prop T^2$. 
Thus, for any $T > 100$K, the cooling time is much shorter than the 
dynamical time $\Omega^{-1} \sim 50\Myr$.}.
We thus expect a $Q<\Qc$ unstable disk to fragment 
into bound clumps encompassing a certain fraction $\alpha$ of the disk mass. 
Another fraction of the disk mass is expected to be in transient arm-like 
density perturbations, partly due to non-axisymmetric modes that could
be unstable even for $Q>1$.  
If the turbulence dissipation somehow becomes slower than the dynamical time,
e.g., because of fragmentation and star formation, 
the balance shifts from bound clumps to transient arms. 
In particular, the old stellar component, in which $\sigma$ does not decay, 
is expected to be part of the transient features. 

Indeed, our simulations (\se{simu}) confirm the visual impression from 
the observed ``chain" and ``clump-cluster" galaxies
that many of the disk clumps have well-defined round boundaries
and are gravitationally bound physical entities.
The simulations also show that the clumps are embedded in a perturbed
disk with elongated transient features, and that a significant
fraction of the clump mass is being exchanged with the surrounding arms
and disk due to tidal effects. 
The notion that the clumps are long lived while the arms are transients
is supported by the finding that the high-redshift star-forming regions 
in spiral arms are younger than the stellar populations in the clumps
of clump-cluster and chain galaxies \citep{elmegreen09}.

\subsection{Disk Fraction}

We adopt as our basic parameter the fraction of mass in the disk 
component within the characteristic radius of the disk $\Rd$,
\be
\f \equiv \frac{\Md}{\Mt} \, ,
\label{eq:f}
\ee
where the total mass $\Mt$ within $\Rd$ includes the contributions of the disk
and the spheroid of dark matter and stars.
The maximum possible value of $\f$ is $\beta$, the fraction of baryons 
including disk and bulge within the disk radius,
\be
\f \leq \beta \equiv \frac{\Mb}{\Mt} \, .
\label{eq:beta2}
\ee
The ratio of disk to total baryonic mass is then
$\Md/\Mb = \beta^{-1}\f$, so that a bulge-less disk is $\f = \beta$,
and a disk-less bulge is $\f =0$.

Expressing the circular velocity at $\Rd$ as
\be
(\Omega \Rd)^2 = V^2 \simeq \frac{G\Mt}{\Rd} \, ,
\label{eq:circ}
\ee
and using the approximate relation 
\be
\Md \simeq \pi \Rd^2 \Sigma \, ,
\ee
we obtain a simple expression for $Q$:
\be
\Qc \gsim  Q \simeq \sqrt{3} \f^{-1} \frac{\sigma_r}{V} 
\, ,
\label{eq:Q}
\ee
where we have assumed an isotropic three-dimensional velocity dispersion  
$\sigma = \sqrt{3}\sigma_r$.
The initial clump radius becomes 
\be
\Rc \simeq \frac{\pi}{6} \f \Rd \, ,
\label{eq:Rc}
\ee
so the typical clump mass is
\be
\Mc \simeq \frac{\pi^2}{36} \f^2 \Md \, .
\label{eq:Mc}
\ee

 
It would be worthwhile evaluating the likely range of the disk fraction
$\f$ and its maximum value $\beta$ in the SFGs.
We note up-front that the observed clumpy disks, with $Q \sim 1$,
rule out naked disks, $\f=1$. 
This would have required an extremely high dispersion,
$\sigma \sim V$, more compatible with a spheroid.
It would have also implied that the clumps are more massive than the observed
clumps.
Indeed, $\beta$ is guaranteed to be less than unity due to the dark-matter 
contribution to the mass within the disk radius.

What is the typical value of the baryonic fraction $\beta$?
While $\beta \simeq 0.5$ in the Milky Way today,
current estimates for SFGs at $z \sim 2$ range from 0.5 to 0.8
\citep{bournaud08,forster09}.
One way to crudely estimate $\beta$ from a theoretical perspective 
is by relating the disk radius to the halo virial radius $\Rv$ via 
\be
\Rd \simeq \lambda \Rv \, , 
\label{eq:lambda}
\ee
where $\lambda$ is roughly the halo spin parameter \citep{fe80,mmw98}. 
The average value of $\lambda$ from
cosmological $N$-body simulations and tidal-torque theory
is $\simeq 0.04$ and rather independent of mass and time 
\citep[e.g.,][]{bullock01_j}. On the other hand,
the observed typical radii of $\Rd \sim 10\kpc$ for
the high-redshift clumpy disks 
and the indicated virial radii of $\Rv \simeq 100 \kpc$
for the halos of these galaxies 
imply an effective value of $\lambda \simeq 0.1$ for these 
galaxies \citep{bouche07,genzel08} (see a discussion of this excessive
disk sizes in \se{size}).
Assuming further, very crudely, that the halo is an isothermal sphere of 
virial mass $\Mv$ hosting a baryonic mass $\Mb$ of gas and stars,
we can approximate 
$\Mt \simeq \lambda \Mv + \Mb$, 
and obtain
\be
\beta \simeq \frac{\fb}{\fb+\lambda} \, ,
\label{eq:beta}
\ee
where $\fb\equiv \Mb/\Mv$ is the baryonic fraction within the virial radius.
The value of $\fb$ could be lower than the universal value of 
$\simeq 0.16$ because of mass loss, e.g., due to supernova-driven winds 
\citep{ds86}.
For $\lambda \sim \fb \sim 0.1$ we obtain $\beta \simeq 0.5$, and
with $\lambda$ and $\fb$ in the ranges $0.04-0.1$ and $0.05-0.15$,
respectively, the value of $\beta$ ranges from $0.33$ to $0.75$.
Replacing the crude isothermal sphere with a more realistic NFW profile of
a low concentration parameter $C \simeq 4$, as appropriate for halos of the
relevant masses at $z \sim 2$ \citep{bullock01_c}, we obtain 
inside $0.1\Rv$ a value of $\beta \simeq 0.63$.
The observational indications and this crude theoretical estimate
lead us to adopt $\beta = 0.6$ as our fiducial value.

\Equ{Q} with $Q \sim 1$ implies that disk fractions in the range 
$\f\sim 0.3-0.6$ are consistent with the observed range of velocity 
dispersions $\sigma_r/V \sim 0.17-0.35$ \citep{genzel08}.
For such values of $\f$ in \equ{Mc}, the typical disk clumps are predicted
to involve a few percent of the disk mass each, as observed
\citep{elmegreen07,genzel08}.

If each clump has contracted by a factor of two from its initial radius
into virial equilibrium,
the internal one-dimensional velocity dispersion within the clump, 
given by $\sigc^2 \simeq (1/3) G\Mc/(0.5\Rc)$,  
can be written using \equ{Rc} and \equm{Mc} as
\be
\frac{\sigc}{V} \simeq \frac{\sqrt{\pi}}{3} \f \, .
\label{eq:sint}
\ee
Since the radial velocity dispersion of the clumps relative to each other, 
based on \equ{Q}, obeys $\sigma_r/V \simeq \f Q/\sqrt{3}$, we conclude that
$\sigc \simeq \sigma_r$ when $Q \sim 1$.

\secpush
\section{Self-Regulated Clumpy Disk}
\label{sec:dispersion}

For a disk of a given surface density in a given potential well,
namely given $\Sigma$ and $\Omega$, the instability provides a feedback 
loop that can drive an unstable disk into a self-regulated, 
marginally unstable
state, where $Q \sim \Qc$ and the clumps are as massive as they could be.
This means that the velocity dispersion $\sigma_r$ is kept at the maximum
possible value for which $Q$ is still below the critical value for rapid 
instability.
This self-regulation is achieved if the instability and fragmentation process 
itself is the generator of velocity dispersion, and if it is capable
of doing so on a timescale comparable to the timescale of turbulence decay.
Then, if $\sigma_r$ is temporarily low such that $Q$ falls below $\Qc$, 
the fragmentation becomes more efficient and it drives $Q$ up toward $\Qc$.
If $\sigma_r$ overshoots to high values such that $Q>\Qc$, the disk becomes 
stable, the fragmentation process is suppressed, and $\sigma$ is allowed to
drop such that $Q$ settles to a value just below $\Qc$. 

The turbulence of a gaseous disk dissipates by shocks and by a
turbulent cascade to the viscous scale.
With a turbulence energy of $E\sim (3/2)M\sigma_r^2$,
and for a dissipation rate of $\dot E \sim M \sigma_r^3 /\Re$
where $\Re$ is the radius of the largest eddy,
the dissipation timescale is $\tdis \sim (3/2)\Re/\sigma_r$.
If $\Re \simeq \Rc$, and we denote the disk dynamical crossing time by
\be
\td \equiv \Omega^{-1} = \frac{\Rd}{V} \, ,
\label{eq:td}
\ee
we obtain from \equ{Rc}
\be
\tdis \simeq 1.4\, Q^{-1}\, \td \, .
\label{eq:tdis}
\ee
For $Q$ of order unity, the dissipation timescale is on the order of the 
disk dynamical time, with no explicit dependence on $\alpha$ or $\f$.
In order to maintain a self-regulated marginally unstable state, 
the stirring up of turbulence must occur on a comparable timescale.


In an unstable disk with $Q<1$, 
the density perturbations grow as $e^{|\omega| t}$,
where $\omega$ obeys the dispersion relation \citep[e.g.,][eq. 6.66]{bt08}
\be
\omega^2 = \kappa^2 - 2\pi\, G\, \Sigma\, |k| + \sigma_r^2\, k^2 \, .
\label{eq:omega}
\ee
One can see that the wave number of the fastest growing mode is 
$k=\pi G \Sigma/\sigma_r^2$, and the growth rate of this mode is 
given by $\omega^2 = \kappa^2 -(\pi G \Sigma/\sigma_r)^2$.
This growth rate, which vanishes for $Q=1$, becomes comparable to the dynamical
time for values of $Q$ slightly below unity, e.g., $|\omega| = \Omega$ 
for $Q = 1/\sqrt{2}$ (assuming $\kappa=\sqrt{3}\Omega$). 
The system should maintain a value of $Q$ in this ballpark slightly below unity
in order for the fragmentation to react in time to variations in 
the turbulence level. 

What is the mechanism that stirs up the disk and maintains the required 
velocity dispersion?  
Some turbulence is generated by feedback from stars and 
supernovae (\se{survive}). Turbulence is also generated by dense gas clumps
that flow in as part of the cold streams, but this process is
not self-regulated and may stabilize the disk (see \se{stab}).
However, the highly perturbed fragmented disk is capable of self-regulating
itself by the gravitational interactions within it, without the help of
external energy sources, and thus maintain the disk in the marginally
unstable state.
One particular mechanism of this sort for generating velocity dispersion 
in the disk is the gravitational encounters between the bound disk clumps
\citep[see][]{wada02,agertz08,tasker09}. 
We demonstrate below that these clump encounters by themselves may be 
sufficient for self-regulating the disk instability.

The timescale for the clump encounters to generate the velocity dispersion
required for a given value of $Q$ can be estimated as follows.
First, given that the disk thickness is roughly $(\sigma_z/V)\Rd$,
with a vertical velocity dispersion $\sigma_z \simeq \sigma/\sqrt{3}$,
the spatial number density of clumps is 
\be
n_{\rm c} \simeq \frac{\alpha \Md/\Mc}{\pi \Rd^3 \sigma_z/V} \, ,
\label{eq:nc}
\ee
where $\alpha$ is the fraction of the disk mass that is in clumps at a given
time. 
Second, the cross section for encounters that involve an energy change 
of order $\sigma^2$ can be estimated 
by\footnote{If the clump internal velocity dispersion
$\sigma_{\rm int}$ was smaller than 
the clump-clump velocity dispersion $\sigma$, the cross section for the most
effective individual encounter was actually $\pi\Rc^2$, which is larger by
a factor $(\sigma/\sigma_{\rm int})^4$, 
but the energy change in each encounter was smaller by a similar factor.
Therefore, \equ{cross} is indeed 
the effective cross section for a total change of $\sim \sigma^2$ in a
series of such encounters.}
\be
\tilde{\sigma} \simeq \pi (G \Mc /\sigma^2)^2 \, . 
\label{eq:cross}
\ee
%
Then, the characteristic time for the encounters to generate a specific energy 
change $\sim \sigma^2$  
is estimated by $\te \simeq (n_{\rm c} \tilde{\sigma} \sigma)^{-1}$. 
Using \equs{circ}, 
\equm{Q} and \equm{Mc}, we obtain 
\be
\te \simeq 2.1\, \alpha^{-1}\, Q^4\, \td \, .
\label{eq:tc}
\ee
Recall that this expression is valid for $Q$ of order unity, but note the 
strong dependence of $\te$ on slight deviations of $Q$ from unity,
with no explicit dependence on $\f$.
The kinetic energy of the gaseous clumps could be dissipated
in head-on collisions, but the cross section for such collisions is
smaller by a geometrical factor of order a few than the cross section for   
effective gravitational encounters.
We therefore expect the net effect of clump encounters
to be of stirring up the dispersion on
the timescale estimated in \equ{tc}.

Self-regulation could be obtained by clump encounters alone if 
$\te \sim \tdis$, i.e., $\alpha^{-1} Q^5\sim 0.67$ when comparing 
\equs{tc} and \equm{tdis}.
This is valid, for example, when $Q \sim 0.67$ and $\alpha \sim 0.2$.
Recall that $Q$ is expected to be slightly below unity in the self-regulated 
unstable state both because $\Qc \simeq 0.68$ is the critical 
value for instability of a thick disk and because such a value of $Q$ allows 
perturbation growth on a dynamical timescale. 
Independently, preliminary observations and simulations
\citep{elmegreen05,elmegreen07,ceverino09}
indicate that the value of $\alpha$ can range from 10\% to 40\%.
A value of $\alpha$ in this ballpark could indeed be expected if 
the unstable wavelength along the tangential direction is comparable to
the radial wavelength $\lambda_{\rm c}$, and if clumps form at the 
regions of positive interference of the tangential and radial waves.
We thus adopt $Q \simeq 0.67$ and $\alpha \simeq 0.2$ as our fiducial case.

If the dissipation timescale somehow becomes significantly 
longer 
than the 
disk dynamical time, the balance between dispersion generation by clump
encounters and turbulence decay is obtained over a timescale longer than 
$\td$.  Such a slowdown of the dissipation rate may result from the
development of inhomogeneities and fragmentation of the disk gas, through
the increase in size of the largest eddies and the possible decrease in
cross section for eddy collisions.
It may also happen when the stellar component of the disk
becomes substantial, as $\tdis$ is inversely proportional to the gas fraction
in the disk. 
In this case of slower dissipation the self-regulation
in the marginally unstable state can be achieved by clump encounters with 
a slightly larger value of $Q$ and/or a lower value of $\alpha$.

The gaseous component between the clumps
is expected to participate in the instability and fragmentation
process and thus to share the stirred-up clump velocity dispersion.
This is especially true for the large-scale transient features,
which participate in the stirring up process as they are sheared and
stirred up themselves. 
The inclusion of such additional stirring-up effects
may be analogous to increasing $\alpha$ in the analysis of clump encounters
above.

\secpush
\section{Clump Migration and Disk Evacuation}
\label{sec:mig}

The same clump interactions that generate velocity dispersion, 
and the dynamical friction exerted by 
the rest of the disk on the giant clumps, also affect the systematic 
rotation velocity of the clumps and make them migrate into the center. 
Both the smooth and clumpy components of the disk are participating in 
this process, so $\alpha$ should be 
replaced by unity for the purpose of estimating the migration time.
Since the disk just outside a given clump has systematically
lower angular velocity, it drains angular momentum from the clump.
Over a timescale of $\alpha \te$, it would systematically reduce
the angular velocity of the clump by $\sigma$ and therefore 
cause it to migrate on a timescale $\alpha (V/\sigma) \te$.
However, the inner disk pushes the clump systematically outward 
on a similar timescale . The net effect acts on a timescale longer
by $(V/\sigma)$ than that of each side of the disk alone
\citep{goldreich80,ward97}
The migration time is thus
\be
\tm \simeq \alpha \left(\frac{V}{\sigma}\right)^2 \te
\simeq  2.1\, Q^2 \f^{-2} \td \, .
\label{eq:tm}
\ee
Note that the migration is rapid when $Q$ is low and when $\f$ is high,
namely when the disk is massive and unstable, but the migration timescale
does not depend explicitly on $\alpha$.

Given that only a fraction $\alpha$ of the disk is in the giant clumps,
the evacuation rate by clump migration of a mass comparable to the 
entire disk mass is
\be
\Med \simeq \frac{\alpha\Md}{\tm} \, ,
\label{eq:Med}
\ee
and the timescale for the evacuation of the entire disk mass is
\be
\tevac \simeq \alpha^{-1} \tm 
\simeq 10.5\, \alpha_{.2}^{-1}\, Q^2\, \f^{-2} \td \, ,
\label{eq:tevac}
\ee
where $\alpha_{.2} \equiv \alpha/0.2$.
With $Q=0.67$ and a dominant disk of $\f = 0.35$, say,
we have $\tm \simeq 7.6\,\td$. 
Then with $\alpha=0.2$, the evacuation time is $\tevac \simeq 38\,\td$.
With $\td \simeq 50\Myr$, 
we obtain $\tm \simeq 380\Myr$ and $\tevac \simeq 1.9\Gyr$.
These estimates are consistent with the findings in 
simulations of isolated disks \citep{bournaud07c}.
When $\f$ is slightly smaller, the timescales are somewhat longer, consistent
with observational estimates \citep{elmegreen05}.

The angular-momentum transfer and mass flow in the perturbed disk involve 
several other processes beyond the clump-clump interactions and dynamical 
friction. For example, the clumps induce transfer of disk angular momentum 
outward, causing much of the disk to accrete inward at a rate that is
comparable to the mass inflow rate directly associated with the clump 
migration itself \citep{sari04}. This accretion rate 
exceeds the migration rate by $d/h \gsim 1$, where $d$ is the distance 
along the disk at which density waves shock and dissipate and $h$ is the 
vertical scale height of the disk. 

The sheared transient features, which are also present in the unstable disk,
exert torques which also cause angular-momentum transfer from the inside out. 
The associated mass inflow rate, based on \citet{shakura73}, is
\be
\Mdshear \simeq 3^{3/2}\, G^{-1} \sigma_r^{3} Q^{-1} \tilde\alpha \, ,
\ee
where $\tilde\alpha$ is the dimensionless angular-momentum flux density.
According to \citet{gammie01}, it is given by 
\be
\tilde\alpha = [(9/4)\tilde{\gamma}(\tilde{\gamma}-1)\Omega\tdis]^{-1} \, ,
\ee
where $\tilde{\gamma}$ is the adiabatic index in the equation of state.
For $\tilde{\gamma}=5/3$, appropriate for a turbulent medium,
and with $\Omega\tdis \simeq 3$ providing the maximum effect,
we have $\tilde{\alpha} \simeq 0.13$.
The timescale for evacuating the entire disk by the inflow associated
with the transient features becomes
\be
\tshear \simeq 7.5\, Q^{-2} \f^{-2} \td \, .
\label{eq:trans}
\ee
With $Q \simeq 0.67$, this is three times longer than the disk evacuation
time by clump migration, \equ{tevac}, but with $Q$ closer to unity the 
two evacuation times become comparable. 

We thus conclude that by considering only the clump migration, ignoring the 
other processes causing mass inflow, we obtain an order-of-magnitude
estimate for the disk evacuation rate that can serve as  
a lower limit to the actual evacuation rate.

\secpush
\section{Cosmological Input Rate by Streams}
\label{sec:accretion}

The key to understanding the high abundance of massive disks with giant clumps
at high redshift, as well as the high SFR in them,
is the continuous vigorous input of gas into these disks via cold streams.
The average relative accretion rate into halos of mass $M$ in the standard
$\Lambda$CDM cosmology is given to a good approximation by 
\be
\frac{\dot M}{M} \simeq 0.47\, (1+z)_3^{2.25}\, M_{12}^{0.15}\, \Gyr^{-1} \, ,
\label{eq:Mdot}
\ee
where $(1+z)_3\equiv (1+z)/3$ and $M_{12}\equiv \Mv/10^{12}\msun$.
This has been derived by \citet{neistein06} as a useful fit to an
analytic prediction based on the EPS approximation and it has been shown
to fit well the halo growth rate measured in the Millennium cosmological 
N-body simulation \citep{neistein08a,genel08}.   
\citet{dekel09} have shown using cosmological hydrodynamical simulations that
at high redshift
the same expression approximates the average relative baryonic input rate 
by cold streams into the disks of galaxies of a given baryonic 
mass, $\Mbd/\Mb$. The actual accretion rates into individual galaxies
range from three times lower than the average to twice the average.
A fraction $\gamma$ of the stream mass is in clumps
of mass larger than $0.1\Mb$, with an average of $\gamma \simeq 0.33$
and an effective range from $\gamma \ll 1$ to $\gamma$ as high as $\simeq 0.6$
(we denote $\gamma_{.33}\equiv \gamma/0.33$).
The rest is in a smoother form consisting of mini-minor gas-dominated clumps 
and smooth gas in unknown proportions because of limited resolution in the
simulations. 

We wish to relate this accretion rate to the disk dynamical crossing time,
which can be written as $\td = 48\, t_{48} \Myr$.
A dynamical time of $48 \Myr$ is obtained for a disk radius $\Rd =10 \kpc$ and
a rotation velocity $V = 200 \kms$, which are the typical values of $\sim
10^{11}\msun$ SFGs at $z \sim 2$. 
Then, \equ{Mdot} defines the accretion time,
\be
\ta \equiv \frac{\Mb}{\Mbd} \simeq 44\,\tau^{-1}(z,M)\,\td \, ,
\label{eq:ta}
\ee
where the explicit redshift dependence and the weak mass dependence 
are in the factor   
\be
\tau(z,M) = t_{48}\, (1+z)_3^{2.25}\, M_{12}^{0.15} \, .
\label{eq:tau}
\ee
Note that $\tau \sim 1$ for the typical SFGs at $z \sim 2$.
As long as the relevant SFG disks are observed to
have similar radii and velocities at different redshifts, 
namely similar dynamical times, 
\equ{tau} is useful for describing the redshift dependence of $\ta$.

Alternatively, it could be useful to use $\Rd = \lambda \Rv$, 
as in \equ{lambda}, 
if $\lambda$ is rather constant in some redshift range,
as deduced from cosmological simulations and tidal-torque theory
\citep[e.g.,][]{bullock01_j}. 
If the characteristic disk rotation velocity is comparable
to the virial velocity, we have 
$\td \simeq \lambda\, \tv$,
where $\tv \equiv \Rv/\Vv$ is the halo virial crossing time.
The cosmological relation between halo virial radius and virial
velocity\footnote{For the standard $\Lambda$CDM cosmology ($\omm=0.28$,
$\oml=0.72$, $h=0.7$), in its Einstein-deSitter regime, $z>1$, 
the standard virial radius and velocity are related 
to the virial mass as $\Rv \simeq 308\kpc\, (1+z)^{-1} M_{12}^{1/3}$ and
$\Vv \simeq 118\kms (1+z)^{1/2} M_{12}^{1/3}$.
Thus $V_{200} \simeq R_{100}\, (1+z)_3^{3/2}$, where
the quantities are in units of $200\kms$ and $100\kpc$, respectively.
The virial crossing time is roughly a constant fraction of the Hubble time,
$\tv \simeq 0.15\, \th$.
In this regime the inverse of the expansion factor can be approximated by
$(1+z) \simeq 6.6\, \th^{-2/3}$, where the Hubble time $\th$ is in Gigayears.
}
implies that $\tau$ of \equ{tau} can be alternatively expressed as
\be
\tau(z,M) = \lambda_{.1}\, (1+z)_3^{0.75}\, M_{12}^{0.15} \, ,
\label{eq:tau2}
\ee
where $\lambda_{.1} \equiv \lambda/0.1$.
Note that if $\lambda$ is constant in time, as opposed to $\td$ being constant,
the redshift dependence of $\ta$ is weaker.
The two expressions for $\tau$, \equs{tau} and \equm{tau2}, 
coincide at $z \simeq 2$
for $t_{48}=\lambda_{.1}=1$. 
Based on the observed SFGs, it seems that the disk dynamical time is
roughly constant at lower redshifts $z \leq 2$. On the other hand,
a constant $\lambda$ may be more sensible from a theoretical point 
of view at $z \geq 2$ 
(see a discussion of disk sizes in \se{size} below). 

At $z=2$, when the cosmological time is $\th \simeq 3.25 \Gyr$, the virial
crossing time is $\tv \simeq 0.48 \Gyr$ and $\ta \simeq 2.1 \Gyr$ 
compared to the disk dynamical time of $\td \simeq 48 \Myr$.

\secpush
\section{Cosmological Steady State}
\label{sec:SS}

\subsection{Evolution Equation}

The high number density of massive SFGs 
that appear to be rotating disks broken into giant clumps
indicates that they constitute a large fraction of the galaxies in 
halos of $\sim 10^{12}\msun$ at $z \sim 2$. 
This implies that the presence of giant clumps is not a short episode in the
galaxy lifetime that disappears in one migration time, on the order of 
$400 \Myr$, but rather a long-term phenomenon that lasts for a Hubble time, 
on the order of $3 \Gyr$ or more.
The key for a long-lived disk with giant clumps is, again, the continuous, 
intensive gas supply into the disk by cold and rather smooth streams, 
valid at high redshift \citep{dekel09}.

We wish to work out how $\f$ evolves in time, as the disk is being built and
drained and as the bulge grows.
The time derivative $\dot\f$ can be derived from
\be
\f = \beta\, \frac{\Md}{\Mb} \, . 
\label{eq:fss}
\ee
The baryon growth rate $\Mbd$ is the total baryonic accretion rate from 
\equ{ta},
and the rate of change of disk mass, $\Mdd$, is
the difference between input to the disk and output from it,
\be
\Mdd \simeq (1-\gamma)\,\Mbd - \Med(\f) \, ,
\label{eq:Mdd}
\ee
with $\Med(\f)$ from \equs{tm} and \equm{Med}.
For completeness, the associated growth rate of the bulge is
\be
\Msd \simeq \gamma\,\Mbd + \Med(\f) \, .
\label{eq:Msd}
\ee

The factor $\gamma$ allows for the possibility that only the smooth fraction 
$(1-\gamma)$ of the incoming streams adds to the disk mass, while the more 
massive clumps along the streams merge into the spheroid independent of the 
disk instability; that is, a standard merger. 
One assumes here that, on their way to join the bulge,
those massive external clumps do not spend a significant
amount of time as proper disk members (with a circular velocity $V$ and a 
dispersion $\sigma$) compared to the rest of the disk material. 
They therefore do not increase the disk surface density
and thus do not speed up the migration and evacuation of the in situ clumps.
This assumption is justified because the massive external clumps are several
times more massive than the internal disk clumps, and their effective masses
are even larger because they are likely to carry with them 
dark matter, so their inward migration time by dynamical friction is much 
shorter than the disk evacuation time.
From the simulation analyzed by \citet{dekel09}, we learn that
the average value of $\gamma \simeq 0.33$ 
refers to external clumps more massive than $0.1\Mb$.
This could perhaps serve as a crude estimate for the dividing line between
``massive" stream clumps that lead to major or minor mergers without 
joining the 
disk, and the smoother component involving ``mini-minor" clumps that do join 
the disk and spend a non-negligible time there.
We estimate further, based on the simulation, that
about a third of the galaxies are fed by smooth streams of 
$\gamma \ll 1$, and another third have an excess of massive incoming clumps,
$\gamma \gsim 0.5$.

As a simple example, we work out the case $\beta={\rm const.}$,
following the notion that the virial baryonic fraction $\fb$ and the
effective spin parameter $\lambda$ that enter \equ{beta} are constant in
time.
The gas input rate is assumed to follow the cosmological mean for that mass
and redshift.
We also assume for simplicity that $\Mv$ remains in the broad range
$10^{11}-10^{12.5}\msun$, taking advantage of the very weak mass 
dependence of the accretion time and the dynamical time.
Using the definitions \equs{tevac} and \equm{ta} in \equ{Mdd},  
we obtain for the rate of change of $\f$
\be 
\dot\f \simeq \beta\,(1-\gamma-\beta^{-1}\f)\,\ta^{-1} -\f\, \tevac^{-1}(\f)\, ,
\label{eq:dotf}
\ee
in which the $\f$-dependence is explicitly specified.
We intend to solve this differential equation for $\f(t)$.

\subsection{Steady State Attractor} 
\label{sec:ss}

We first note that
for a given accretion rate, there should be a steady-state configuration
where $\dot\f=0$; that is, $\f$ remains fixed at a critical value, $\fss$.
This configuration
serves as an ``attractor": if $\f > \fss$, the efficient disk
evacuation and spheroid growth make $\f$ decline toward $\fss$, and if
$\f < \fss$, the efficient streaming into the disk forces $\f$ up 
back to $\fss$.
In order to estimate the value of $\fss$, we set $\dot\f$ to zero in
\equ{dotf}, namely 
\be
\frac{\ta}{\tevac(\fss)} \simeq \beta\, (1-\gamma)\, \fss^{-1} -1 \, .
\label{eq:fss2}
\ee
The relevant time ratio, from \equs{tevac} and \equm{ta}, is  
\be
\frac{\ta}{\tevac} \simeq 4.18\, \alpha_{.2}\, Q^{-2}\, \tau^{-1}\, \f^2 \, ,
\label{eq:tate}
\ee
so, the equation is a depressed cubic polynomial equation for $\fss$,
\be
\fss^3 +b\, \fss -c =0 \, ,
\label{eq:fss3}
\ee
\be
b\equiv 0.24\, \alpha_{.2}^{-1}\, Q^2\, \tau \, , \quad
c\equiv b\, \beta\, (1-\gamma) \, .
\label{eq:fss30}
\ee
The solution (e.g., by Cardano's method) is
\be
\fss =u-\frac{b}{3u} \, ,
\label{eq:fss31}
\ee
\be
u=\left( \frac{c}{2} \right)^{1/3} 
\left[ 1 +\left( 1+\frac{4b^3}{27c^2} \right)^{1/2} \right]^{1/3} \, .
\label{eq:fss32}
\ee
With the fiducial values $Q=0.67$, $\alpha=0.2$ and $\beta=0.6$, 
and with $\gamma$ in the range $0-0.33$, we obtain at $z=2$, 
\be
\fss \simeq 0.32-0.25 \, .
\label{eq:fss4}
\ee
Simple solutions of \equ{fss3} in certain limits are discussed in 
\se{app_ss}. 

The redshift dependence of $\fss$ is rather weak. For example, with $\gamma=0$,
we obtain $\fss \simeq 0.38$ and $0.25$ at $z=9$ and $1$, respectively
(where we have assumed $\lambda \simeq 0.1$ at $z\geq 2$ 
and $\td \simeq 48 \Myr$ at $z\leq 2$).
The mass dependence is even weaker.
Varying $\beta$ in the range $0.5-0.75$ yields $\fss \simeq 0.28-0.35$.
Varying $\alpha$ in the range $0.1-0.4$ gives $\fss=0.37-0.26$.
We see that when varying the parameters within their likely ranges 
about the fiducial values, the value of $\fss$ at $z \sim 2$
remains quite stable about $0.3$. 
Thus, the steady-state configuration is expected to be
with a bulge mass comparable to the disk mass, $\f \simeq \beta/2$,
to within a factor of two.

\subsection{Convergence to Steady State}

\begin{figure}
\vskip 9.0cm
\includegraphics{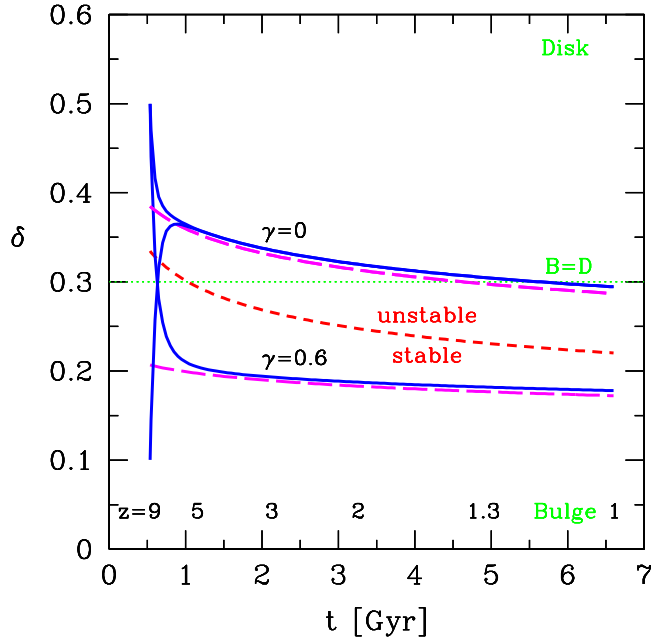}
\caption{Steady state and bimodality.
The solid blue curves show the evolution of the disk-to-total mass ratio 
$\f$, according to \equ{dotf}, 
for the fiducial case ($Q=0.67$, $\alpha=0.2$, $\beta=0.6$, $\td=0.015\,t$
and an average cosmological inflow rate).
Shown are two extreme cases in terms of the presence of massive clumps
in the incoming streams: very smooth streams ($\gamma=0$) and very
clumpy streams involving mergers ($\gamma=0.6$).
\Equ{dotf} is solved starting at $z=9$ with either a dominant 
disk ($\f=0.5$) or a dominant bulge ($\f=0.1$). 
Independent of the initial configuration, as long as the disk is unstable 
the system converges within less than a Hubble time to a quasi-steady
state where $\f$ is slightly above $\fss$ (long-dashed magenta curves). 
The critical value for stabilization by incoming stream clumps, 
$\fs$ of \equ{fs}, is shown as a dashed red curve
for $\gamc=0.33$ and $\vin=1.2$.
It implies that streams that are smoother than average ($\gamma < 0.33$)
produce a long phase of an unstable giant-clump disk with a comparable bulge
that can last till after $z\sim 1$.
Streams that are more clumpy than average ($\gamma > 0.33$) 
stabilize the disk and lead to a bulge-dominated system with a less massive, 
smoother disk.
Therefore, the actual solution for $\gamma = 0.6$ deviates from the lower 
curve shown.
}
\label{fig:tcs}
\end{figure}

With the average accretion rate evolving on a timescale that is not
much longer than the disk evacuation timescale, we need to verify that the
system indeed approaches a quasi-steady state,
and find out how far it might be from $\fss$ where $\dot\f=0$.
\Fig{tcs} shows numerical solutions for $\f(t)$ of \equ{dotf} 
with \equ{tate}, for the fiducial choice of parameters
and $\td=0.015\,t$, \equ{tau2}.
The initial conditions are set, as an example, at the reionization epoch,
$z_{\rm in} \simeq 9$ or $t_{\rm in}\simeq 0.54\Gyr$. We choose the starting
configuration to be alternatively disk dominated, $\f_{\rm in} = 0.5$,
or bulge dominated, $\f_{\rm in} \ll 1$.
Independent of the initial conditions, the system settles
within a Hubble time to a value of $\f$ near $\fss$, which is also shown. 
This can be understood via the approximate solutions to \equ{dotf} discussed
in \se{app_ss}. 
The actual steady-state value of $\f$ is
slightly higher than $\fss$ because the latter declines with time
on a timescale that is somewhat larger than the timescale for changes in $\f$,
driven in this range by $\tevac$.

We conclude that throughout the broad redshift range of interest
the system tends to be in a steady state in which the bulge
and disk masses are comparable, to within a factor of two.
The disk fraction in this steady state is larger in cases where 
the fraction of massive clumps in the incoming streams is smaller.
Note that the evolution curves shown in \fig{tcs} are derived, so far, 
under the assumption that the disk is always unstable to the formation 
of giant clumps.
We will see next that this assumption may be valid only for streams that are
smoother than average, namely a low $\gamma$ and then a high $\f$.
An excess of incoming steam clumps may stir up the disk to stabilization
with a low value of $\f$.

The observed high-redshift SFGs indeed seem to indicate convergence
into a steady state with a bulge of mass comparable to the disk mass, 
$\f \sim 0.3$, as predicted here.
\citet[][and private communication]{genzel08} have measured for six SFGs 
the ``concentration" $C$, defined
as the ratio of total dynamical masses within the radii of 0.4" and 1.2",
which roughly correspond to $(1/3)\Rd$ and $\Rd$, respectively. 
Allowing for 20\% of the mass within the inner radius to come from dark matter 
and disk (appropriate for an NFW halo profile and a uniform disk profile),
the observed value of $0.8 C$ could be interpreted
as the bulge-to-total mass ratio within the disk radius, $0.8 C\simeq\beta-\f$. 
When they plot in Fig.~5 the value of $C$ against the time $t_*$ of onset 
of star formation
in that galaxy\footnote{$t_*$ is the time to form the current stellar mass 
at a constant SFR that equals the current SFR --- a model that fits 
the data well \citep{forster09}}, 
they find that $C$ is rising 
monotonically with $t_*$ for $t_* < 0.5 \Gyr$ until it settles 
to a rather constant value near $C \simeq 0.38$ for $t_* \simeq 0.5-2\Gyr$. 
This is consistent with the quick convergence to a steady state as predicted
in \fig{tcs}, and with the steady-state configuration of $\f \simeq 0.3$ 
(for $\beta=0.6$ and $\gamma<0.33$), which corresponds to a bulge-to-total 
ratio $0.8C \simeq 0.3$.

\secpush
\section{Stabilization by External Stream Clumps}
\label{sec:stab}

Beyond the self-regulated stirring up of the disk by the in situ disk clumps 
themselves (\se{dispersion}), the incoming streams provide an alternative 
source for generating velocity dispersion in the disk. 
Unlike the stirring by the internal products of the disk instability, which
is naturally self-regulated, the external source could in principle
push $\sigma$ up to a level that stabilizes the disk.

When a smooth stream of gas density $\rho_{\rm s}$
hits a disk of much higher gas density $\rho_{\rm d}$, conservation
of momentum implies that the kinetic energy  
added to the disk is roughly a fraction $\rho_{\rm s}/\rho_{\rm d}$ of the total
kinetic energy of the stream, while most of the incoming energy 
turns into heat that is radiated away.
This implies that the smooth component of the streams, which is less dense 
than the disk by more than an order of magnitude, 
is not effective in stirring up the disk.

However, the streams contain a clumpy component with gas densities comparable
to the disk density. In the impact of such an external dense clump with the 
disk, about half the kinetic energy is expected to turn into turbulence in 
the disk.
The timescale for stirring up the disk to a given value of $\sigma$ 
by external gas clumps that stream in with a velocity $\Vin$
can thus be estimated by
\be
\ts \simeq \frac{\Md \sigma^2}{(\gamc/2) \Mbd \Vin^2} \, .
\label{eq:ts}
\ee
The assumption here is that incoming clumps that constitute a fraction
$\gamc$ of the stream mass share most of their kinetic energy and angular 
momentum with the disk.  
This $\gamc$ is not necessarily the same as the $\gamma$ used in \se{SS}
to compute the evolution of an unstable system into a steady state. 
It may include the same massive clumps, under the assumption that they 
stir up the disk before they end up merging with the bulge, as well as  
mini-minor clumps (e.g., less massive than $0.1\Mb$), which do not merge 
quickly with the bulge but are dense enough for stirring up the disk. 
We defer a more accurate estimate of the fraction of dense clumps in the 
streams to a future work based on simulations with higher resolution.
Meanwhile, for the purpose of the current simplified analysis,
we adopt $\gamc = 0.33$, the average value deduced 
from the MareNostrum simulation used in \citet{dekel09}\footnote{It 
may be interesting to note that \citet{khochfar09} 
estimate from a semi-analytic model that $\simeq 18\%$ of the stream kinetic 
energy is required for driving a velocity dispersion as observed, 
consistent with the $\gamc/2$ factor predicted by our theoretical argument,
with our average value $\gamc \simeq 0.33$.}.
We mark $\vin\equiv\Vin/\Vv$, and read from the same simulation that
on average $\vin^2 \sim 2$.

We see from \equs{tm} and \equm{tevac} that 
$\sigma^2/V^2 \simeq \te/\tevac$, so with $\ta$ from \equ{ta}, 
\equ{ts} becomes
\bea
\ts &\simeq& 2\, \gamc^{-1} \vin^{-2} \beta^{-1} \frac{\ta}{\tevac(\f)}\,
             \f\, \te \\
    &\simeq& 88\, \gamc^{-1}\, \vin^{-2} {\beta}^{-1}\, \tau^{-1}\, Q^2\, 
             \f^3\, \td\, . 
\label{eq:ts2}
\eea
The first equality tells us that when the external clump accretion rate is
sufficiently higher than the disk evacuation rate, 
the stirring by external stream clumps is more effective than the stirring
by internal disk clumps, which operates on a timescale $\sim \te$. 
This generation of turbulence by an external source is not self-regulated 
and could stabilize the disk if it operates on a timescale that allows it to 
balance the dissipation of turbulence, namely if $\ts \leq \tdis$.
Using \equs{tdis} and \equm{ts2}, this condition translates to stabilization
by external stream clumps once  
\be
\f < \fs \simeq 0.25\, (\gamc\, \vin^2\, \beta\, \tau)^{1/3}\, Q^{-1} \, . 
\label{eq:fs}
\ee
Naturally, the disk tends to be stabilized by the streams when the streams 
contain more dense clumps and the accretion rate is high,
and especially when $\f$ is low due a significant contribution from 
the spheroid.
The latter is because a low $\f$ requires a low critical $\sigma$ for 
$Q \simeq \Qc$, which makes it easier for a given stream to drive $\sigma$ 
above the critical value.
 
Note that the mass dependence of $\fs$, entering through $\tau$ in \equ{fs},
is very weak. For a halo mass smaller by an order of magnitude, $\fs$ is
smaller by $\sim 10\%$.
The redshift dependence entering through $\tau$ is somewhat more pronounced,
but it is still rather weak; it depends on whether $\td$ or
$\lambda$ are assumed to be constant in time, \equs{tau} or \equm{tau2},
namely $\fs \prop (1+z)^{0.75}$ or $(1+z)^{0.25}$, respectively.

\Fig{tcs} shows $\fs$ for the fiducial case $\gamc =0.33$ and $\vin=1.2$.
We learn that for average clumpiness, $\gamma \simeq \gamc \simeq 0.33$, 
the value of $\fs$ happens to be in the ballpark of the steady-state value 
of $\f$. 
The value of $\gamma \simeq 0.33$ thus marks the transition between two 
very different evolution tracks.
In the cases of streams smoother than average, $\gamma < 0.33$, 
the stirring by external stream clumps is unimportant, 
as $\f > \fs$ at all times.
This is especially true for the case $\gamma \ll 1$ shown in the figure.
In this case, and as long as the accretion rate follows the cosmological
average, the disk is in a long-term giant-clump phase near steady state 
with $\f$ slightly above $\fss$. For $\gamma \ll 1$, the value of  
$\f$ ranges from $\f \simeq 0.4$ at very high $z$ to $\f \simeq 0.3$ 
at $z\sim 1$, 
namely the bulge contains between a third and a half of the baryonic mass. 
This is compatible with the bulges detected in typical SFGs at high redshift,
as mentioned at the end of \se{SS}. 

In the cases where the streams are more clumpy than average, $\gamma > 0.33$, 
the stabilization by external stream clumps
becomes important, as $\f < \fs$ at all times.
The disk does not develop a giant-clump phase (and if it somehow did at
very high redshift, it was a short episode). In such a case, the bulge 
grows predominantly by the merging incoming massive clumps. 
The system does not evolve according to the evolution track shown in \fig{tcs}
for $\gamma=0.6$.
Instead, it settles to either $\f \simeq \fs$ or $\f=\beta(1-\gamma)$, 
whichever is smaller.
In this case the bulge is a high fraction of the baryonic mass,
on the order of $\gamma$.
 
Our analysis thus predicts a robust {\it bimodality\,} in the galaxy population
already at high redshift, driven by the spread in the degree of clumpiness
in the streams feeding the galaxies.
A small deviation of $\gamma$ from the average value $\simeq 0.33$ triggers
a qualitative difference in the evolution pattern, even if the overall mass
growth rate follows the cosmological mean.
This bimodality is consistent with the observed high abundance of clumpy 
star-forming, extended disks at $z \sim 2$, alongside with the realization
that a significant fraction of the massive galaxies are dominated by compact 
spheroids and show low SFR already at these redshifts
\citep{kriek06,dokkum08}. We predict that many of these high-redshift spheroids 
are surrounded by smooth, stable disks containing about a third 
of the baryonic mass, which continue to evolve secularly and form 
stars quiescently (\se{stellar}).
Such secondary disks are indeed apparent in some of the HST NIC2 images of 
compact passive spheroidal galaxies at $z \sim 2-3$ 
\citep[][Fig.~1]{dokkum08}. 
This configuration, of a dominant compact stellar spheroid and a smooth 
secondary disk, is not very different from the predicted
remnants of certain binary wet major mergers
\citep{springel05,robertson08,governato09},
not to be confused with the extended, high-redshift disks of giant clumps 
forming stars that are not dominated by the central spheroid 
\citep{bournaud09}.

The evolution described in \fig{tcs} assumes continuous gas input at the 
average cosmological rate and a constant level of clumpiness in the streams.
Time variations in the stream properties can shift a galaxy from
an unstable to a stable mode and vice versa. 
An example is illustrated in the simulated evolution of a Milky-Way-type galaxy
termed ``Via Lactea" \citet[][Fig.~5]{agertz09}, where the overall growth 
rate drops to below-average
levels between $z \simeq 2.5$ and $1.8$. As a result, the bulge-to-disk ratio
grows and $\f$ drops below $\fs$, bringing the unstable giant-clump
phase to an end. When the accretion resumes at $z \sim 1.8$, 
it takes the system a Hubble time, about $4\Gyr$, to rebuild a disk that 
may bring it back to the unstable regime with $\f > \fs$ after $z \simeq 1$. 
By that time, the disk is already in the star-dominated regime and may never
resume instability (\se{stellar}).

\secpush
\section{Star Formation}
\label{sec:SFR}

The observed SFGs of $\Mb \sim 10^{11} \msun$ at $z \sim 2$ show that the 
giant disk clumps are the sites of intense star formation.
This indicates that the formation of clumps is a necessary condition for 
the main mode of star formation in high-redshift galaxies. 
While the physics of this mode of star formation is beyond the scope of 
the current paper, we discuss here certain implications of our analysis of
gas streams and disk instability, combined with the observed overall SFR, on
the process of star formation in the clumps.  

The observed overall SFR of $\sim 100 \Msy$ across the disk
\citep{elmegreen07,genzel08} is only slightly smaller than 
the theoretically predicted gas input rate via cold steams into the 
central galaxies of dark-matter halos that are comparable in abundance 
to these SFGs \citep{dekel09}. This leads to the
approximation that the overall SFR in the disk
is a significant constant fraction of the gas accretion rate from \equ{ta}: 
$\Mstdd \simeq s\, \Mbd$ with $s \sim 0.5$.
The stars formed earlier in the incoming external clumps are not part of 
the star formation in the in situ disk clumps.

As a simple model for the SFR in the disk clumps, we assume that
the clump mass $\Mc=\Mg+\Mst$ remains constant as it turns gas into stars, 
and that the SFR in the clump is $\Mstdc=\Mg/\tst$ with a constant $\tst$.
Then $\Mstdc(t) = (\Mc/\tst) \exp(-t/\tst)$.
This means that most of the stars form during a period $\sim \tst$
with a SFR on the order of $\Mc/\tst$.
The fact that a significant fraction of the accreted gas is observed to form
stars in the clumps while they are still in the disk implies that $\tst$ is not
much longer than the clump migration time $\tm$.
On the other hand, the observations indicate that the number of star-forming 
big clumps, $\sim 5$, is comparable to the number of clumps seen in 
simulations (\se{simu}) and consistent with our fiducial values of 
$\Mc$ and $\alpha$ from the instability analysis. 
If the clumps survive for a migration time (but see \se{rad}), 
this implies that $\tst$ is not much smaller than $\tm$.
We conclude that $\tst \sim \tm$.

This is consistent with the cosmological steady state (\se{ss}) because the
requirement that the SFR be comparable to the accretion rate 
is equivalent to the approximate condition for steady state, $\tevac \sim \ta$.
Indeed, if we write $\tst=\alpha \Md/\Mstdd$, we obtain  
\bea
\tst &\simeq& \alpha\, \beta^{-1} s^{-1} \f\, \ta \\
     &\simeq& 8.8\,\alpha_{.2}\,\beta^{-1} s^{-1} \tau^{-1}\f\,\td \, ,
\label{eq:tst}
\eea
which is comparable to $\tm$ from \equ{tm} for our fiducial unstable disk
with $\f=0.3$ at $z \sim 2$, where $\tst \simeq 8.8\, \td \simeq 420 \Myr$. 

Based on the above simple model, the gas fraction in the clump when it
coalesces with the bulge is $\fg = \exp(-\tm/\tst) \simeq e^{-1}$.
This implies that
the disk clumps merge into the central spheroid while they
are still relatively gas rich.
The process is thus similar to a wet major merger,
where the gas shocks, heats up, cools, and condenses to the center, leading to
a compact spheroid significantly smaller than the disk size
\citep[e.g.,][]{dc06,covington08}.
This may explain the compactness of observed spheroids at $z \sim 2$
\citep{dokkum08}.
In the late phases of spheroid growth, when $\f$ is smaller,  
$\tm/\tst$ is larger, so
the mergers of the disk clumps into the spheroid are less dissipative, 
making the late-formed spheroids less compact.

The SFR efficiency in the disk clumps, based on \equ{tst} and with 
$\tc \simeq 0.5 \td$, is
\bea
\eta &\equiv& \frac{\Mstdc}{\Mc/\tc} = \frac{\tc}{\tst} \\ 
 &\simeq& 0.06\, \alpha_{.2}^{-1} \beta\, s\, \tau\, \f^{-1} \, .
\eea
For the fiducial case, with $\f \simeq 0.3$
at $z \simeq 2$, we obtain 
$\eta \simeq 0.06$.
Similar high efficiencies were estimated for $z\sim 2$ SFGs based on the 
pressure in the star-forming clumps \citep{lehnert09}.
However, such values of $\eta$ are higher 
than the efficiencies in the most efficient star-forming clouds
at low redshift. The common wisdom, based on the empirical Kennicutt-Schmidt
law \citep{kennicutt89} and on star-formation theory \citep{krumholz09},
is that $\eta$ is of order 1\% in all star-forming environments, independent
of the actual density.
This indicates that the giant disk clumps may not form stars uniformly over 
their whole volume. Instead, stars form with a standard efficiency $\sim 1\%$
in subclumps where the density is two orders of magnitude higher so the 
local dynamical time is an order of magnitude shorter.
This is an observable prediction,
to be addressed when the star-forming clumps are resolved.

\section{Clump Survival and Disruption}
\label{sec:survive}

The appearance of giant gas clumps forming stars in the disks of many SFGs
indicates that they survive for at least several dynamical times.
The same is evident in simulations (see \se{simu}).
One wishes to understand the mechanisms that prevent the clumps from collapsing
into themselves on a free-fall time and help them survive the various
processes that work to disrupt them.
In particular, once the disk clumps form stars at high rates, they could be 
affected by stellar feedback, in the form of supernovae-driven winds or 
radiative feedback from massive stars, which we briefly comment on below.
A preliminary discussion of the possible gravitational origin for the
turbulence pressure supporting the clumps is provided in \se{pressure}.

\subsection{Supernova Feedback}
\label{sec:sn}

Could supernova feedback provide the pressure
support against free-fall collapse?
If so, is there a risk that supernova-driven winds
may remove the remaining gas and disrupt the clumps before they complete their
migration? \citet{ds86} evaluated 
the maximum energy fed into the interstellar gas by supernovae,
taking into account the radiative loses, 
$E_{\rm SN} \simeq \epsilon\, \nu\, \Mstd\, t_{\rm rad}$,
where $\Mstd$ refers to the SFR in each clump,
$\epsilon$ is the energy released by a typical
supernova ($\sim 10^{51} {\rm erg}$), and $\nu$ is the number of supernovae per
unit mass of forming stars (which for a typical IMF is $\nu \sim 1$ per 
$50\msun$).  The characteristic time $t_{\rm rad}$ marks the end of the 
``adiabatic" phase and the onset of the ``radiative" phase of a typical 
supernova remnant, by which it has radiated away a significant fraction 
of its energy and became ineffective. They found that it takes a similar 
time for the expanding supernova shells to reach a significant mutual overlap, 
which allows an even distribution
of the supernova energy (minus the radiative loses) over most of the gas. 
A necessary condition for the clump to be significantly affected by 
supernova feedback is that $E_{\rm SN}$ exceeds 
the binding energy of the clump, $\sim (1/2) \Mc \sigma^2$.
This occurs if the potential well associated with the clump 
is shallow enough, with a three-dimensional velocity dispersion
\be
\sigma^2 \leq V_{\rm SN}^2 
= 2\, \epsilon\, \nu\, \eta\, \frac{t_{\rm rad}}{\td} \, .
\ee
It turns out that in the relevant temperature range the cooling rate
scales approximately as $\Lambda \propto T^{-1}$, which implies that the
ratio of timescales is roughly $t_{\rm rad}/\td \sim 0.01$ independent of
gas density and other clump parameters. 
\citet{ds86} then assumed maximum efficiency in a burst of star formation,
$\eta \simeq 1$, and obtained a critical velocity of 
$V_{\rm SN} \simeq 100 \kms$
as an upper limit for the virial velocity of a system in which one may expect
substantial gas heating or removal by supernova feedback.

With the estimated SFR efficiency of $\eta \sim 0.1$ in each clump as a whole,
the critical velocity is reduced to $V_{\rm SN} \sim 30\kms$. 
This implies that supernova feedback cannot be effective in the 
giant clumps during the steady-state instability phase of a $V \sim 200\kms$ 
disk, where $\f \sim 0.3$ and therefore $\sigma \sim 40-60\kms$ by \equ{Q}. 
The clumps may become more vulnerable to  
supernova feedback, $\sigma < 30 \kms$, once $\f < 0.2$, namely when
the bulge becomes twice as massive as the disk 
\citep[see also simulations by][]{tasker08}. 
However, when the clump fragments, the supernova-driven winds 
may escape harmlessly via low density ``chimneys" through a  
porous interstellar medium \citep[e.g.,][]{ceverino+k09},
and thus not lead to disruption even in clumps of $\sigma$ well below $30\kms$.

\subsection{Radiation Pressure}
\label{sec:rad}

While supernova feedback may be unimportant in the giant clumps,
radiative feedback from O stars may be more effective.
\citet{murray09} estimate that the clumps should be disrupted by  
radiative pressure once about $20\%$ of their gas has turned into
stars, expelling the remaining 80\% of the gas back to the disk.
If true, we can deduce that
the system would maintain a steady state similar to the one
described in \se{SS}, with the disk evacuation by migration replaced by 
an ``evacuation" into dense star clusters that do not participate
in the disk instability any longer. 
The two steady states are similar because the timescales
for clump migration and for star formation in a clump mass are comparable 
(\se{SFR}).
The expelled gas, combined with the gas streaming in from the outside,
would then help keeping the disk gaseous and unstable.

However, if each clump is disrupted this way, its mass becomes
smaller by a factor of $\sim 5$, so the bulge buildup by migration 
becomes slower by a factor of $\sim 5^2$. This is because
the migration time is inversely proportional to the clump mass,
and the mass added to the bulge by each coalescing clump is proportional
to the clump mass. 
This implies that most spheroids have to be made by mergers,
in possible conflict with the cosmological merger rate
and with most semi-analytic simulations of galaxy formation 
(also discussed in \se{conc}).
On the other hand, 
this scenario predicts the common existence of massive thick stellar 
disks, constructed from the slowly migrating star clusters, which may 
be associated with today's thick disks and S0 galaxies.

The stellar populations in the high-redshift clumps can provide an 
observational test that could clarify the survivability of the clumps. 
With $\eta \sim 0.06$, one fifth of the clump mass becomes stars in about 
three clump dynamical times, namely $\sim 75\Myr$. If radiative feedback
suppresses star formation after that time, the spread of stellar ages 
in each clump should not exceed $\sim 50-100\Myr$. The slow migration
implies that the single-age population in the clumps could be as old as 
$\sim 2 \Gyr$.
In contrast, if the gas clumps survive for a migration timescale,
the age spread could be $\sim 0.5\Gyr$, and giant clumps should not show
populations older than $\sim 1 \Gyr$. In this case, we predict that
the typical giant clumps should be gas rich and forming stars at a high rate
while they also contain a stellar population that has been formed in a similar
rate during the preceding few Gigayears. There is observational evidence in
favor of the latter \citep{forster09}
Furthermore, if the clumps survive for a migration timescale, one expects
most of the clumps at large disk radii to contain stellar populations 
younger than $\sim 0.5\Gyr$, while the clumps at smaller radii may also contain 
older populations.
In contrast, if the clumps have lost their gas reservoir in less than 
$\sim 100\Myr$, they must have formed all their stars near the radius where 
they were formed and where they orbit for more than $1 \Gyr$, thus showing 
no obvious age gradient with radius.
Finally,
the massive radiation-driven outflows from clumps should be observable
as systematic blueshifts at the clump locations. A preliminary search for
such outflows yielded a null result (K. Shapiro, private communication).

\secpush
\section{Galaxy Size}
\label{sec:size}

As mentioned in \se{intro}, the observed SFG disks of $\Md \sim 10^{11}\msun$ 
at $z \sim 2$ are surprisingly
extended, with radii $\sim 10\kpc$, comparable to the Milky Way today.
These disk radii indicate $\lambda \sim 0.1$ in \equ{lambda} 
whereas the average value is expected to be smaller by a factor 
of $2-3$ \citep{bullock01_j}.
We note that
the proposed scenario of stream-driven unstable disks provides several 
possible explanations for the extended disk sizes, 
as well as for the compact spheroids.

The mass inflow in the disk into its center, 
by clump migration due to dynamical friction and encounters
or by torques involving the transient shear and tidal
features (\se{mig}), is associated with angular momentum transfer into the 
outer disk, which tends to stretch it further out.
If the baryons in the bulge have lost all their angular momentum
to the disk, by the time the bulge mass is comparable to the
disk mass, the angular momentum per unit mass in the outer disk
should be about twice the average, so the disk is expected to be 
twice as extended as it would have otherwise been.
The timescale for reaching this state, \equs{tevac} and \equm{trans},
is on the order of one or two orbital times, namely a few hundred Megayears,
and the configuration of comparable disk and bulge, $\f \simeq 0.3$,
is indeed the typical configuration expected in the clumpy steady state
(\se{SS}).

Furthermore, in massive high-$z$ disks that are fed by smooth streams, 
the generation of clumps and therefore the regions of intense star formation
are expected to extend to the outer disk.
First, the streams that determine the angular momentum of the disk
tend to be coherent and to join the disk with an impact parameter
comparable to the disk radius.
They therefore generate high gas surface density $\Sigma$ at large radii,
as opposed to the exponential profiles of today's disks. 
The relative smoothness and low density of the high-$z$ streams make them 
inefficient drivers of turbulence (\se{stab}), 
so they are not expected to push $\sigma$ in the outer disk to high values.
Finally, with a near flat rotation curve driven by the dark halo and the
rather uniform disk outside the bulge, the angular velocity declines with
radius roughly as $\Omega \prop r^{-1}$.
These three factors allow $Q\prop \sigma\Omega/\Sigma$ to become low
and admit instability at large radii. This is seen in simulated
galaxies \citep{ceverino09} and is consistent with the appearance of a broad,
large-radius ring of giant clumps in some of the SFGs \citep{genzel08}.

The situation is very different in low-redshift disks.
Unlike the halos at $z \sim 2$ that are fed by intense, coherent, narrow 
streams, a $10^{12}\msun$ halo at low redshift is fed from all directions 
by rather slow and erratic accretion \citep{keres05,db06,ocvirk08,dekel09}.
The exponential profile tends to provide a sufficiently high $\Sigma$ only at 
small radii, where the halo-driven rotation curve is closer to a
solid-body rotation with a constant $\Omega$, so the instability is
not preferred at large radii.



Another possibly relevant mechanism is supernova feedback from early star 
formation
in the central regions of galaxies. It is expected to be more effective 
at high redshift, when both the gas density and the SFR were higher.
Supernova-driven winds may preferentially remove low angular momentum gas 
that has settled at earlier times in smaller radii, and thus leave behind 
a more extended system with higher angular momentum 
\citep{md02,governato07,scannapieco08}.

Observations by \citet{law09} indicate that among the high-redshift galaxies
with baryonic masses of a few$\times 10^{10}\msun$, slightly below the range
of SFGs addressed in our paper, about a third seem to be supported by velocity
dispersion rather than rotation, with $\sigma/V \sim 1/2-1$.
These galaxies tend to be gas rich (up to 90\%) and not overly extended,
with typical radii $R\sim 1-2 \kpc$.
Such a galaxy could be formed if fed by cold flows
of low impact parameters that carry little angular momentum.
Alternatively,
if these flows are overly clumpy, they could be effective in stirring up
an excessively high dispersion in the gaseous galaxy. 
The absence of a substantial, unstable disk component in such a galaxy during 
its history may explain the relatively low past SFR and thus the observed 
high gas fraction.

\secpush
\section{Gas \& Stellar Disks at High \& Low Redshifts}
\label{sec:stellar}

The high-redshift disks are observed to be gas rich, at the level of 50\%
and perhaps more \citep{bouche07,daddi08}. 
In contrast, today's massive disks are dominated by stars, with gas fractions 
of $\sim 20\%$ or less. 
Whereas both the gas and stellar components of the disk participate in the
disk instability through their contributions to the local self-gravity, 
the gas dissipation introduces a long-term difference between the two 
components. 
While the decay of gas turbulence and gas cooling 
allow the velocity dispersion or sound speed to decline
even when the system is self-regulated and in a cosmological steady state,
the stars tend to maintain the velocity dispersion that they have acquired 
during their history. The young stars may still have a $\sigma$ 
that is not much larger than that of the gas, 
but the older stars are likely to have a higher $\sigma$. 
Therefore, the young stars actively participate in driving the perturbation 
growth and they may even follow the gas into the bound clumps, while the 
older stars tend to form transient perturbations and eventually join the 
stable component that just adds to the external potential well and thus 
helps stabilizing the disk.

The cosmological gas accretion rate plays an important role in determining
the gas fraction in the disk.  
If stars form only in massive gas clumps, the disks are expected to remain 
gas rich as long as the gaseous input into the disk manages to 
replenish the disk 
mass on a timescale shorter than the time for the entire disk to turn into 
clumps, which is comparable to $\tevac$, namely 
\be
1 < \frac{(1-\gamma)\Mbd}{\Med} \simeq 
0.24\,\beta\, (1-\gamma)\, \alpha_{.2}^{-1}\, Q^{2}\, \f^{-3}\, \tau \, .
\label{eq:gaseous}
\ee
The second equality is based on \equs{tevac} and \equm{ta}. 
For the unstable disk in steady state (\se{ss}), with $\gamma < 0.33$,
this ratio is $\sim 2-3$ at $z \sim 2$, and it varies slowly with redshift.
This implies that the disk should remain gas rich as long as the input 
is dominated by gas, which is more likely at high redshift.
This indicates that the analysis performed in the current paper so far is 
a useful approximation for the high-$z$ SFGs.
At $z <1.5$, the penetration of cold streams into the centers 
of halos more massive than $10^{12}\msun$ becomes limited 
\citep{db06,ocvirk08,keres09}, 
so the input rate of cold gas to the disks is lower than implied by \equ{Mdot}.
At these late epochs, the fraction of stars in the accreting matter becomes
higher.
The disks gradually become dominated by the stars, in clumps and in transient 
features or spread out in the disk by shear and tidal stripping.

A more accurate analysis of disk instability, especially at low redshift,
should therefore deal with multi-component disks.
The axisymmetric instability of a two-component disk has been studied by 
\citet{jog84} and \citet{rafikov01}.
Denoting the velocity dispersions of stars and gas 
$\sigs$ and $\sigg$, respectively (with the latter standing for the speed of
sound if thermal pressure dominates), and defining $\Qs$ and $\Qg$
following \equ{Q} separately for each component, the effective $Q$ relevant
for the instability of the combined system is approximately
\be
Q^{-1} = 2\,\Qs^{-1} \frac{q}{1+q^2} 
            + 2\,\Qg^{-1} \frac{\siggs q}{1+\siggs^2 q^2} \, ,
\ee
where $\siggs \equiv \sigg/\sigs$ and $q$ is the dimensionless wave number
$q \equiv k \sigs/\kappa$.
The first term has to be slightly modified to take into account the
dissipationless nature of the stars\footnote{for 
a dissipationless component, the $q$
dependence in the first term should be replaced by 
$q^{-1}[1-\exp{(-q^2)}I_0(q^2)]$, 
where $I_0$ is the Bessel function of order 0.}, 
but this correction makes only a small difference \citep{rafikov01}.
The system is unstable once $Q<1$. 
The most unstable wavelength corresponds to the $q$ that minimizes $Q$;
it lies between $q=1$ for $\siggs =1$ and $q \simeq \siggs^{-1}$ for 
$\siggs \ll 1$.
Note that with $\sigs>\sigg$, the stellar disk by itself may tend to
be less unstable than the gas disk by itself, $\Qs>\Qg$, but through its
contribution to the self-gravity that drives the instability, the stellar disk
can help the gas component de-stabilize the disk. The combined system can 
be unstable for axisymmetric perturbations
even when each of the components has a $Q$ value above unity. 

If $\siggs=1$, the two components can be treated as one in the instability 
analysis.  In order to illustrate the effect of 
different values of $siggs$, consider, for example, the  
case of equal mass densities for the gas and stars, $\Sigg=\Sigs$.
When $\siggs=1$, the criterion for instability $Q<1$ translates
to $\Qg<2$ (or $\Qs<2$).
If however the stars are ``hotter", $\siggs=0.5$ say,
the criterion for instability (obtained at $q \simeq 1.6$) becomes $\Qg < 1.42$ 
(or $\Qs <2.84$), so the error made in $\Qc$ by ignoring the higher 
velocity dispersion of the stars (namely 2 versus 1.6) is about 30\%.  
In this case, the instability driven by the cooler gas component is only 
slightly affected by the ``hotter" stellar component.
When the stellar disk is much ``hotter" than the cold gas, $\siggs \ll 1$, 
the instability criterion becomes $\Qg <1$, so the stars
become part of the stabilizing component of the system.

In the solar neighborhood, \citep[according to][\S 6.2.3]{bt08}, 
the gas fraction is $\sim 25\%$, the stars are much ``hotter" than the gas,
$\siggs \simeq 0.18$, and the separate $Q$ values are $\Qg \simeq 1.5$ 
and $\Qs \simeq 2.7$. 
This gives a combined value of $Q \simeq 1.2$, 
indicating that the solar neighborhood is stable
for axisymmetric perturbations because of the high velocity dispersion of the
dominant stellar component. However, it is unstable for non-axisymmetric
perturbations, only 20\% variations in the surface
density or the sound speed of the gas can generate axisymmetric instability.   

We note that disk stabilization can be helped by the tendency of
the gas temperature in the smooth, warm disk component not to drop 
significantly below $10^4$K, corresponding to a speed of sound $\sim 15\kms$. 
This is because atomic cooling is ineffective at lower temperatures, 
and because supernova and stellar feedback tend to heat the gas to such 
temperatures \citep{wolfire03}. 
With $V \sim 200 \kms$, such a minimum speed of sound would not allow $Q$ 
to be kept below unity once $\f < 0.13$.
This implies that once the cold disk becomes less than 20\% of the total
baryonic mass, the disk tends to become stable against the formation of
giant clumps. 

Could the $z\sim 2$ thick disks of the SFGs evolve to the thick disks
of today's spiral galaxies?
In the massive disk galaxies at low redshift, 
the thick disk is typically about a quarter of 
the total disk mass, and in smaller disks it could be as high as one 
half of the disk mass \citep{yoachim06}.
If at $z\sim 2$ half the baryonic mass is in the disk ($\f \sim 0.3$) 
that becomes the thick stellar disk of today, and if, say, the total 
baryonic mass has doubled since then and all of it has settled in a new 
thin disk, 
then today's thick disk fraction is expected to be
about a third, marginally consistent with  the observed fractions.
On the other hand, the velocity dispersion in the high-$z$ stellar disks might 
be too high to be consistent with today's thick disks. Many of them may end 
up as S0 galaxies, or as elliptical galaxies if they go through significant
mergers. 
The simulations described in \se{simu} \citep{ceverino09} indeed demonstrate 
that the evolution into an S0-like configuration is likely.
However, it is clear that not all the massive high-redshift SFGs 
end up as S0's today,
because the mean comoving number density of SFGs
($\gsim 10^{-4} \Mpc$) is higher than that of today's S0's ($<10^{-4} \Mpc$).

\secpush
\section{Cosmological Simulations}
\label{sec:simu}

\begin{figure}
\vskip 15.5cm
\includegraphics{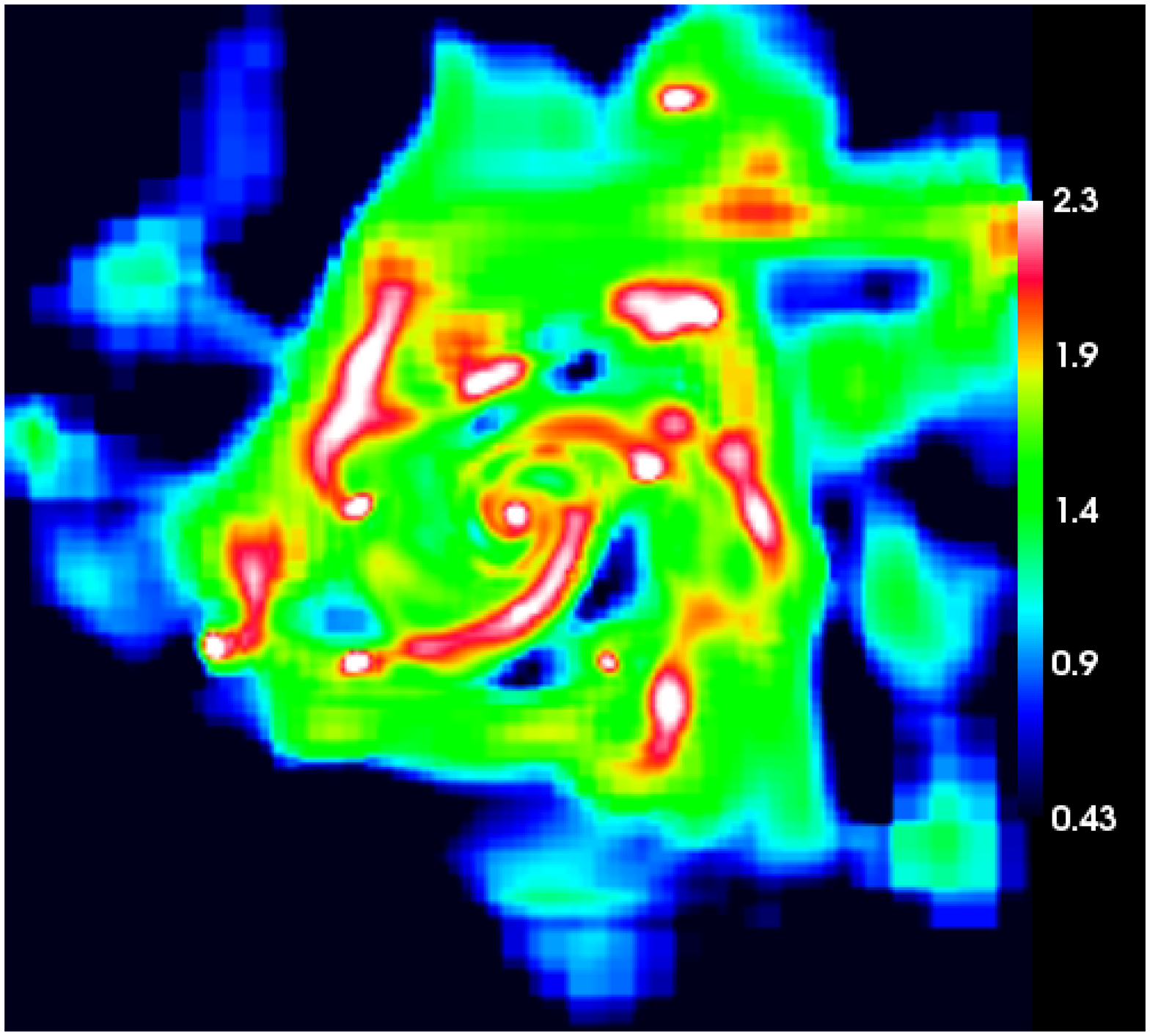}
\includegraphics{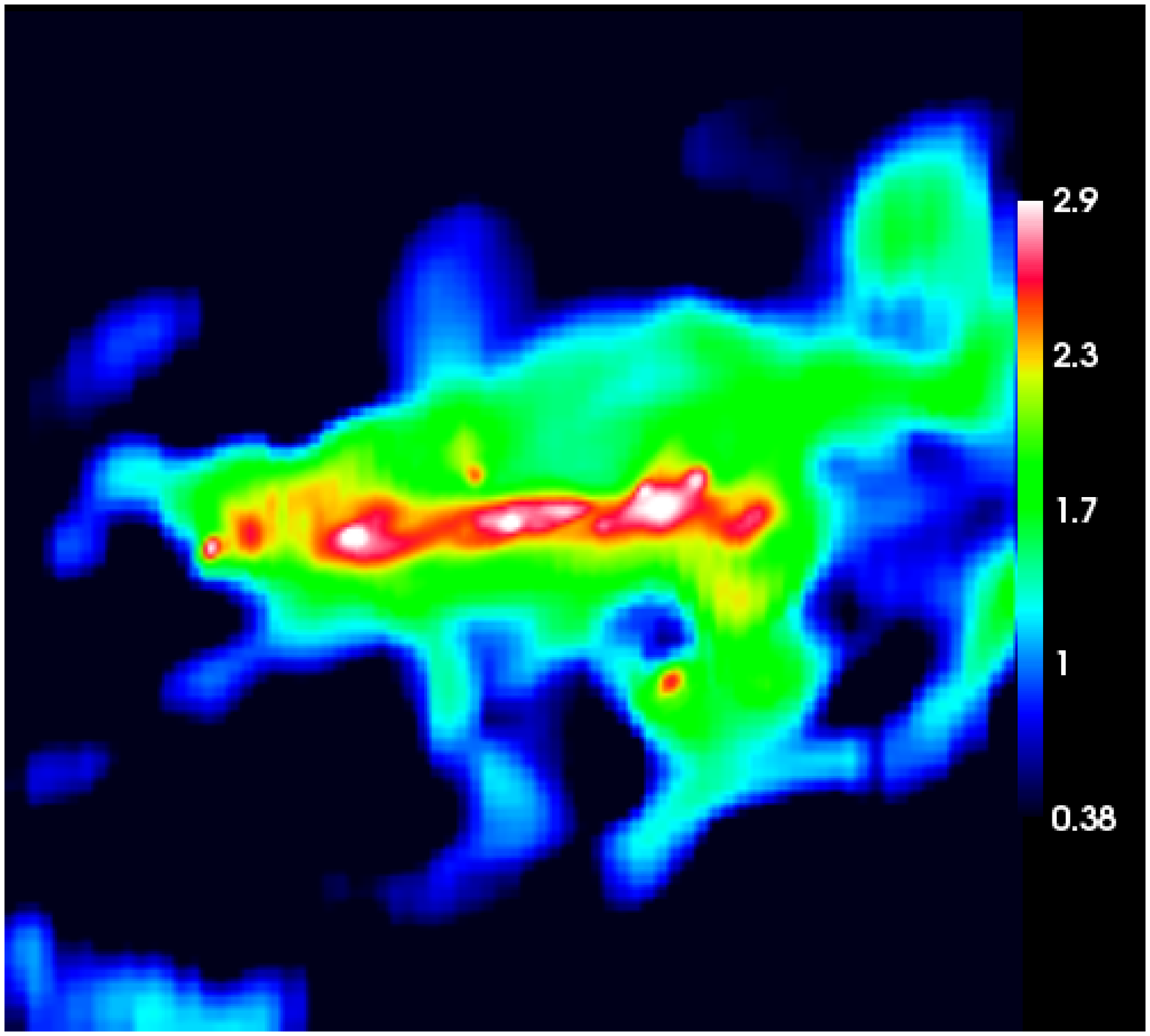}
\caption{Gas surface density of a galaxy at $z \sim 2.3$
from a high-resolution cosmological simulation \citep{ceverino09}.
The box side is $15\kpc$.
The color code is log surface density in units of $\msun \pc^{-2}$.
Only gas with density $n > 0.1\, {\rm cm}^{-3}$ is shown.
The face-on view (top) shows an extended disk broken into several giant 
clumps and sheared perturbations, similar to many observed SFGs.
The edge-on view demonstrates that this is a well defined disk, resembling the
observed ``chain" systems.
The disk is fed by streams, not emphasized in this picture because they are 
largely below the threshold density for the plot.
With $\f \simeq 0.26$, this system is in a steady state at the late stages 
of the clump-instability phase.
This galaxy, in a halo of $4\times 10^{11}\msun$, is smaller than the typical 
SFG, but it resembles the typical appearance of SFGs and it demonstrates 
the general behavior predicted by our theoretical analysis. 
}
\label{fig:MW3}
\end{figure}
 

\Fig{MW3} provides a sneak preview of the results from 
zoom-in cosmological simulations that are analyzed in more detail in a 
companion paper \citep{ceverino09}.  
Shown are the first maps of gas density in a properly resolved galaxy 
at $z \simeq 2.3$, which clearly show a disk broken into giant clumps and
sheared features.
They are brought here for the sake of demonstrating the validity of the 
simple analysis presented above.

These zoom-in cosmological simulations are performed about individual galaxies
using a Hydrodynamic Adaptive Refinement Tree code \citep{krav97,krav03} 
with a spatial resolution better than $70 \pc$ (in physical units). 
The code incorporates the relevant physical processes, including
atomic and molecular gas cooling and photoionization heating,
star formation, metal enrichment, and feedback from stars and supernovae,
as described in \citet{ceverino+k09}.
The unique feature of this code for the purpose of simulating disk instability
is that it allows the gas to cool down to $\sim100$ K. 
This is a key to resolving the turbulence Jeans mass and permitting
the disk fragmentation into giant clumps, not seen before in cosmological
simulations. 

The galaxy shown in \fig{MW3}
is one of a few simulated galaxies, most of which showing 
similar features, although they are on the small side of the massive SFGs 
observed (as they were originally selected for another purpose,
to match a virial mass of $10^{12}\msun$ at $z = 1$).
In the specific case shown, at $z=2.3$, the halo virial parameters are
$\Mv \simeq 4 \times 10^{11}\msun$ and $\Rv \simeq 70 \kpc$. 
The disk extends to $\Rd \simeq 6 \kpc$ with a rather flat rotation curve
of $V \simeq 180\kms$. The disk mass is $\Md \simeq 1.1 \times 10^{10} \msun$,
in comparable fractions of gas and stars.
The stellar spheroid is $\Ms \simeq 1.0 \times 10^{10} \msun$,
and the total mass within $\Rd$ is $\Mt \simeq 4.2 \times 10^{10} \msun$, 
so the disk-to-total ratio is $\f \simeq 0.26$. This value is near the 
predicted steady-state value for $\gamma \lsim 0.33$.

The map clearly reveals a disk configuration dominated by a few giant clumps,
containing a fraction $\alpha \simeq 0.12$ of the disk mass. 
By inspecting the time evolution of this galaxy, we deduce that 
most of the clumps seen in the map were formed in situ in the disk, 
and only two of them came from outside. These two are embedded in dark-matter 
halos while all the other clumps are ``naked" baryonic clumps. 
The edge-on view shows that the clump deviation from the mid-plane is
comparable to and smaller than the intrinsic disk thickness.  
This is both evidence for in situ clump formation and a typical
characteristic of observed ``chain" galaxies \citep{elmegreen06}.
With the bulge comparable to the disk, with the stellar
component already quite developed, 
and with a non-negligible clumpiness of $\gamma \simeq 0.2$ in the 
incoming streams,
this galaxy seems to be
at a late stage of its giant-clump phase. 
The value of $Q$ for the combined gas and stellar disk is below unity in
parts of a ring of radius $\sim 5 \kpc$, but it is above unity for the disk
as a whole. 
This may indicate that this system is at a stage where it is 
gradually becoming stable against axisymmetric perturbations by its 
old stellar component.
Indeed, by the end of this simulation at $z = 1.3$, the galaxy becomes
dominated by its stellar spheroid with only a minor, rather stable gas disk. 
Note, however, that the comparison of the linear instability analysis with 
the simulated clumpy disk should be interpreted with care, taking into account
the fact that the linear analysis is valid at the onset of instability while 
the quantities quoted above are measured from the highly perturbed disk.

Convergence tests that are described in detail in \citet{ceverino09}
demonstrate that the in situ clumps are a result of a 
real physical fragmentation process and not a numerical artifact.
In order to prevent artificial fragmentation due to an unresolved Jeans scale,
we implemented in the code a standard procedure that imposes a pressure floor 
(not by means of a temperature floor), which prevents the 
Jeans length from falling below $\lambda_{\rm J}=\Nj\Delta$, where $\Delta$ 
is the cell size of the numerical grid. The standard value is $\Nj=4$ for
simulations that last a dynamical time \citep{truelove97}. 
To verify the minimum value of $\Nj$ that prevents artificial
fragmentation in a disk that evolves for several dynamical times, 
we repeated our simulation for $400 \Myr$ with the same resolution but using 
$\Nj$ values that range from 1 to 10. 
We found that the mass in clumps and the number of clumps converge to 
constant values for $\Nj=7-10$, while there is an overproduction of 
clumps that worsens with decreasing $\Nj$ in the range $\Nj=1-7$. 
This indicates that the clumps represent a real physical phenomenon once 
$\Nj \geq 7$, so we adopt $\Nj=7$ for our default analysis.

\citet{agertz09} have reported independently and immediately after us 
a similarly clumpy disk 
in their own simulation of a galaxy in a cosmological background, using an 
AMR code very similar to ours and with a comparable resolution.
Applying $\Nj=4$ for the pressure floor, they indeed find more and typically
smaller clumps, as expected from our numerical tests. We conclude that
while their simulation may be partly subject to numerical fragmentation, 
they basically detect the same real physical phenomenon that we see in our
simulations. 

\secpush
\section{Summary and Discussion}
\label{sec:conc}

We studied the Toomre instability of high-redshift, gas-dominated, massive,
thick, galactic disks as they are intensely fed by cosmological cold streams.
Defining the disk fraction $\f\equiv \Md/\Mt(\Rd)$, 
an unstable disk {\it self-regulates\,} itself to $Q \simeq \Qc \sim 1$ 
with a velocity dispersion ${\sigma_r}/{V} \simeq \Qc \f /\sqrt{3}$. 
The disk develops transient elongated sheared features, and fragments into 
a few in situ bound massive clumps, each a few percents of the disk mass and
together involving a fraction $\alpha \sim 0.2$ of the disk mass.
The turbulence is largely maintained by the internal gravitational interactions
within the perturbed disk. In particular, 
the encounters between the disk clumps stir up velocity dispersion
on a timescale $\te \simeq 2.1 Q^2 \alpha^{-1} \td$, 
where $\td \equiv \Omega^{-1} \simeq 50\Myr$,
which is sufficient by itself for matching
the natural timescale for turbulence decay, 
$\tdis \simeq 1.4 Q^{-1} \td$, once $Q \simeq 0.67$ and $\alpha \simeq 0.2$.

The same gravitational encounters and dynamical friction make the giant clumps 
{\it migrate\,} 
to the center on a timescale $\tm \simeq 2.1 Q^2 \f^{-2} \td$ and grow a bulge.
The associated evacuation timescale for the entire disk mass is
$\tevac \simeq 10.5 \alpha_{.2}^{-1} Q^2 \f^{-2} \td$,
which is comparable to the timescale for mass inflow due to torques of the 
transient features.

The cosmological streams feed baryons to the galaxy on a timescale
$\ta \simeq 44 \tau^{-1} \td$, with $\tau \simeq 1$ at $z = 2$ (and varying 
from 0.4 to 2.5 between $z\simeq 1$ and $9$). 
The smooth component of the incoming streams, including small clumps,
replenishes the evacuating disk, while the external massive clumps associated 
with the streams merge to the bulge.
If the galaxy is fed by streams that contain less massive clumps than average, 
$\gamma < 0.33$, the system settles into a near {\it steady state\,} 
with $\f \simeq 0.3$, where the input by streams and the transport from 
disk to bulge maintain a constant bulge-to-disk ratio near unity.

In galaxies where the incoming streams are more clumpy than critical,
$\gamma > 0.33$, the external massive clumps merge into a dominant spheroid, 
$\f < 0.3$, and the dense clumps stir-up the turbulence in the disk to 
levels that {\it stabilize\,} the disk.  
The dependence of the instability on the degree of clumpiness in the streams
introduces a {\it bimodality\,} in the galaxy properties starting already at 
$z \geq 3$.
Streams smoother than average lead to extended, unstable disks with giant 
in situ clumps that form stars at a high rate,
while more clumpy streams help building compact, massive bulges and 
stir up the disks to stabilization followed by suppressed SFR, as follows.

In about half the halos of $\sim 10^{12}\msun$ at $z>1$, 
where the streams are relatively smooth,
the rapid accretion at high $z$ leads to a dense, gas-dominated
disk with $\f \lsim \beta \simeq 0.6$. 
Such a disk is wildly unstable; it grows sheared perturbations and
fragments into a few giant clumps, 
each comprising a few percent of the disk mass.
The clump interactions self-regulate the disk in a marginally unstable state,
with $Q \simeq \Qc$.
The clumps migrate inward in a couple of rotation times and form a spheroid.
The combined effects of penetrating streams,
disk evacuation and spheroid growth make $\f$ approach a near steady 
state within one Hubble time. They maintain the giant-clump phase with $\f
\sim 0.3$ and $\sigma_r/V \simeq 0.15-0.3$ for several Gigayears, 
till well after $z \sim 2$.
During this phase, as the accretion rate gradually slows down, 
the bulge-to-disk ratio grows slowly,
the velocity dispersion $\sigma_r/V$ declines accordingly to keep 
$Q \simeq \Qc$,
the clumps get slightly smaller in proportion to $\f^2$,
and the clump migration and disk evacuation slow down by a similar small 
factor.

In the other half of the massive galaxies,
where the clumpy component of the streams is higher than average,
the dense clumps stir up turbulence in the disk at a level that 
can stabilize the disk against the formation of giant clumps once 
the bulge is more massive than the disk.  
Even if the system starts as an unstable disk forming stars at a very high 
redshift, once fed by clumpy streams it settles within a Hubble time 
to a phase of growth where the disk is stabilized. 
In this case, most of the star formation and the most rapid growth of the 
spheroid occurred during the first Gigayear or two.
Once the disk is stable, it does not form new clumps,
and when the remaining disk clumps have consumed most of their gas or 
disappeared, the SFR is substantially suppressed.\footnote{This is yet 
another context where clumpy streams help the quenching
in massive halos where the potential well is deep \citep{db08,khochfar09}.}


The proposed picture has interesting implications on the sizes of galaxies at 
high redshift.  After the system have grown a bulge comparable to the disk,
the disk is expected to be twice as extended as implied by the standard 
spin-parameter argument, because it acquired the angular momentum lost by 
the material that migrated to the bulge.
Another reason for the extended appearance of the disk is that
the coherent streams tend to join the disks in their outer parts.
This makes the generation of clumps and the resulting star formation 
most efficient in an outer ring.
In turn, the typical high-redshift spheroids are expected to be 
compact because the timescales for in situ clump migration, as well as
the timescale for incoming mergers, are comparable to the star formation 
time in the clumps, making the 
coalescence into the bulge highly dissipative.

The disk giant clumps are the cites of intense star formation,
in an overall rate that follows the gas accretion rate of $\sim 100 \sy$
at $z \sim 2$.
The apparent star-formation efficiency in the clumps is $\eta \sim 0.1$
compared to star formation on a dynamical time.
For a local efficiency at a standard level of $\lsim 1\%$, as observed in
star-forming molecular clouds, the star formation has to be confined to
a cuspy core or
subclumps that are denser than their host giant clumps by one to two orders
of magnitudes.

The giant clumps are not expected to be disrupted by supernova feedback,
but a significant fraction of their gas may be expelled back to the disk
by radiative
stellar feedback \citep{murray09}. This would not change the predicted
steady state in a qualitative way, but it would slow down the bulge growth,
leaving mergers as the dominant mechanism for spheroid formation.
The distribution of stellar ages within the individual disk clumps can
provide an observational constraint on the actual level of clump disruption
prior to their coalescence with the central bulge. If the clumps indeed
underwent an effective gas removal, the spread of ages within each clump is 
not expected to exceed $\sim 100 \Myr$, 
and the outflows from the clumps should be detectable observationally. 

If clump disruption is not too effective during their migration,  
the efficient formation of bulges by clump migration in the high-redshift
disks helps reconciling the models of galaxy formation with 
the observed abundance of stellar spheroids.   
The cosmological major-merger rate seems insufficient for the purpose
\citep[e.g.,][]{jogee08,dekel09,bundy09}. Indeed, not having enough mergers
in the simulations, most semi-analytic models of galaxy formation  
had to assume a high rate of bulge growth from disk instabilities
in order to match the observed abundance of spheroids 
\citep{cattaneo08,parry09}.
Our analysis spells out the origin of this instability. 

As proposed by \citet{elmegreen08c}, the massive black holes observed in
spheroids already at high redshift can naturally originate from seed black
holes that have formed by massive-star coalescence in the dense stellar 
clusters at the centers of the giant clumps. These black holes were shown
to migrate with the clumps into the central spheroid, and to reproduce the 
observed black hole to spheroid mass ratio of $\sim 10^{-3}$. 
The AGN feedback associated with the central black hole can add yet another
quenching mechanism \citep{cattaneo09}, 
which also becomes more effective as the bulge grows.

If the properties of the streams feeding a given galaxy vary in time,
and in particular if the degree of clumpiness in them evolves, 
the galaxy may go through transitions from an unstable disk-dominated 
configuration to a stable bulge-dominated state and vice versa. 
However, after $z \sim 2$, the recovery from a bulge-dominated system
back to an unstable disk takes several Gigayear and may never materialize.

A systematic change in the stream properties is expected after $z \sim 1$,
where the cosmological accretion rate becomes slower  and
the smooth cold streams no longer penetrate very effectively through 
the shock-heated media in massive halos of $\sim 10^{12}\msun$ or higher
\citep{keres05,db06,cattaneo06,ocvirk08}. Once the accretion cannot replenish
the disks on a time scale comparable to the timescale for the disks to turn
into clumps and stars, the galaxies become {\it star dominated\,} and 
eventually stable against axisymmetric modes. Supernova and stellar feedback 
can then add to the stabilization of the gas disk.
The common late disks form predominantly in halos below the threshold mass of
$\sim 10^{12}\msun$, and not necessarily by narrow streams
\citep{bd03,binney04,keres05,db06,bdn07}.
They evolve secularly through non-axisymmetric modes of instability
associated with quiescent star formation.
The turbulent high-redshift disks may end up in today's thick disks 
and S0 galaxies, or in ellipticals through mergers.  
The spheroids continue to grow according to the standard scenario;
bulges develop by slow secular evolution in the disks \citep{bureau05,athan08},
and all spheroids, including today's giant ellipticals,
grow by minor and major mergers.  

We conclude that the typical high-redshift massive galaxies are in 
a phase of evolution that does not have a common parallel in low-redshift 
galaxies.
On one hand, the intense and deeply penetrating coherent, cold gas streams 
keep the disk gas rich and thus drive a wild instability with giant clumps 
forming stars at a high rate, and maintain this phase in steady state for 
cosmological times. 
In parallel, a high degree of clumpiness in some of the high-flux streams 
could stabilize the disks against the formation of giant clumps and form
massive spheroids with low SFR already at high redshift.
We note, however, that there are rare cases at low redshift of relatively 
gas-rich galaxies that somewhat resemble the perturbed clumpy appearance of the 
high-redshift SFGs \citep[e.g., NGC 4303,][]{boissier03}.
They may be scaled down versions of the SFGs, possibly unstable to 
axisymmetric modes but with much lower gas densities and SFR, and probably
fed by overly intense accretion compared to the average at low redshift.

The quenching of disk star formation by a dominant stellar bulge,
which can be termed ``morphological quenching",
is addressed and demonstrated using simulations in \citet{martig09}.
It can explain the existence of red-and-dead early-type galaxies in the field,
that is in halos below the critical mass for virial shock heating,
above which the quenching can be explained in other ways involving
termination of gas supply\citep{db06}.
Morphological quenching predicts the presence of non-negligible stable
gas disks in some of the field early-type galaxies.

Preliminary tests indicate that the theoretical framework proposed here 
is in general agreement with the evolution of galaxies in hydrodynamical 
cosmological simulations of appropriate resolution \citep{ceverino09},
as well as with the observations of high-redshift galaxies.
However, the comparison of the linear instability analysis with the evolved, 
nonlinearly perturbed systems, as simulated or as observed, should be
performed with care.
Our current analysis is meant to provide a simple, basic, theoretical 
framework, but the exact numerical values quoted for the parameters, 
averaged over the whole disk, should not be interpreted too literally.
The simulations allow us to predict the observable appearance of the cold
streams that drive galaxy formation at high redshift, either in emission
as Lyman-alpha Blobs \citep{loeb09,goerdt09}, or in absorption as
Lyman-limit systems or Damped Lyman-alpha systems \citep{dekel09}.


\acknowledgments

We acknowledge stimulating discussions with Frederic Bournaud, Andi Burkert,  
Bruce Elmegreen, Reinhard Genzel, Peter Goldreich, Tobias Goerdt, Mark
Krumholz, Doug Lin, Norm Murray and Amiel Sternberg.
This research has been partly supported by an ISF grant, 
by GIF I-895-207.7/2005, by a DIP grant, by a France-Israel Teamwork 
in Sciences, by the Einstein Center at HU,
by NASA ATP NAG5-8218 at UCSC, and by an ERC Starting Grant (RS).

\appendix

\section{Steady State: Approximate Solutions}
\label{sec:app_ss}

In the limit $4b^3/(27c^2)\ll 1$, the expression in \equ{fss32} becomes
$u\simeq c^{1/3}$.
In our fiducial case at $z \simeq 2$, this limit corresponds to
$(1-\gamma)^2\gg 4/27$, namely smooth streams, leading to steady state
with the disk more massive than the bulge.
For $0 \leq \gamma \leq 0.33$ (where $0.33 \geq \fss \geq 0.26$),
this approximation underestimates $\fss$ by a few percent.

In the other limit, $c^2/b^3\ll 1$, there is an approximate solution of the
sort $\fss=(c/b)(1-\epsilon)$ with $\epsilon \ll 1$; to
first order in $\epsilon$ it is
$\epsilon \simeq c^2 b^{-3}/(1+3 c^2 b^{-3})$.
In our fiducial case at $z \simeq 2$, this limit corresponds to
$(1-\gamma)^2\ll 1$, namely clumpy streams, leading to steady state
with a dominant bulge.
This approximation overestimates $\fss$ by a few percent for $\gamma \geq 0.3$
($\fss \leq 0.27$).

The general behavior of the solution of \equ{dotf} can be evaluated
analytically in the fiducial case when noticing that the second term
involving $\ta$ becomes negligible near $\f \simeq \beta(1-\gamma)$, that is
$\simeq 0.33$ for the fiducial case, not far from the steady-state value.
With $\td \propto t$, the simplified equation 
$\dot\f \simeq -\tevac^{-1}\f$ becomes
$\dot\f /\f^{3} \simeq -6.33\, \alpha_{.2} \lambda_{.1}^{-1} t^{-1}$,
and its solution is
\be 
\f^{-2}-\f_{\rm in}^{-2} \simeq 12.7\, \alpha_{.2}\, \lambda_{.1}^{-1}
\ln (t/t_{\rm in}) \, .
\ee 
According to this, with $\f_{\rm in} \simeq 0.5$, the system is expected
to evolve to near $\fss$ by $t \simeq 2t_{\rm in}$, as seen in the exact
solution, \fig{tcs}.

If we assume instead that $\td$ is the same at all times,
the equation is replaced by
$\dot\f/ \f^{3} \simeq - 2.0\, \alpha_{.2} t_{48}^{-1}$
and the solution becomes
\be 
\f^{-2}-\f_{\rm in}^{-2}
\simeq 4.0\, \alpha_{.2}\, t_{48}^{-1}\, (t-t_{\rm in}) \, .
\ee
Again, the system is expected to evolve rapidly to near steady state,
crudely reproducing the behavior in \fig{tcs}.

The term neglected in \equ{dotf} tends to add a negative contribution
to $\dot\f$ for $\f >\beta(1-\gamma)$, and a positive contribution for 
smaller values of $\f$.   
This speeds up the evolution when the system is far form the steady state
value, and it may either speed it up or slow it down near steady state.

\section{Comments on the Origin of Pressure Support}
\label{sec:pressure}

Since the gas cools rapidly to $10^4$K,
the thermal pressure cannot support
a clump of $\sigma_r \simeq 30 \kms$ against gravitational collapse and star
formation on a free-fall time. The required pressure support must be due to
velocity dispersion. Can the gravitational interactions in the perturbed disk
drive the turbulence inside the clumps? We address this issue using simulations
elsewhere, 
and bring only two preliminary considerations here.

The dissipation rate of turbulence inside a rather uniform gas clump is
\be 
\dot{E}_{\rm dis} \simeq \frac{1.2\,\Mc\ \sigc^3}{0.5\,\Rc}
        \simeq 2.4\, \Mc\, \sigc^2\, \td^{-1} \, ,
\label{eq:Edis}
\ee
assuming a radius $0.5\Rc$ for the clump,
and using in the second equality $\Rc/\sigc \simeq \td$ with $\td$ the
disk dynamical time.
The associated timescale is $\tdis \simeq 0.6\Rc/\sigc$, i.e., comparable
to the dynamical time of the clump, $\tc \sim 0.5 \td$.  
In comparison, the rate of gravitational work done on the clump
as it migrates a radial distance $\Rd$ is
\be
\dot{E}_{\rm mig} \simeq \frac{G\Mt\Mc}{\Rd\,\tm}
 \simeq 0.5\, \Mc\, V^2\, Q^{-2}\, \f^2\, \td^{-1} \, .
\label{eq:Emig}
\ee 
The ratio of the two is
${\dot{E}_{\rm dis}}/{\dot{E}_{\rm mig}} \simeq 1.7 \, Q^4$,
which is of order unity for $Q\sim 1$ and 
about a third for our fiducial $Q=0.67$.
This implies that the gravitational power associated with the migration
is in principle enough for balancing the turbulence decay inside the clumps.
The actual mechanism for pumping up the internal energy in the clumps
could in principle be clump encounters, shear and tidal 
interactions with the transient perturbations, yet to be studied in detail.

The decay of turbulence within the clumps may actually be slower than implied
by \equ{Edis} and thus easier to balance. For example, the dissipation
rate naturally slows down 
in proportion to the gas fraction as the gas turns into stars.
The fragmentation to dense subclumps, advocated in \se{SFR}, 
may have a similar effect by itself. If the giant clump fragments
to $N_{\rm sub}$ subclumps of equal mass in which the
gas density is $n_{\rm sub}$, the turbulence decay rate would change in
proportion to the total cross section for subclump collisions, 
\be
\dot{E}_{\rm dis} \prop N_{\rm sub} \frac{R_{\rm sub}^2}{\Rc^2}
\prop \left( \frac{n_{\rm c}}{n_{\rm sub}} \right)^{2/3} N_{\rm sub}^{1/3}\, .
\label{eq:Efrag}
\ee
If $n_{\rm sub}/n_{\rm c} \sim 36$, to allow $\eta \sim 0.01$ in the
subclumps (\se{SFR}), we obtain that the turbulence decay would slow down
as long as $N_{\rm c} < 10^3$. This implies for $10^9\msun$ clumps that
the dissipation timescale would be longer than the clump dynamical time
as long as the actual gas clouds forming stars are bigger than $10^6\msun$.

\bibliographystyle{hapj}
\bibliography{flows}

\end{document}